\documentclass[11pt,a4paper]{article}

\usepackage{amsmath,amssymb}     
\usepackage{color}
\usepackage{graphicx}
\usepackage{subfig}
\usepackage{cite}                
\usepackage{hyperref}            
\usepackage{multirow,makecell}
\usepackage{braket,bm}
\usepackage{cancel}
\RequirePackage{doi}
\usepackage{hyperref}
\usepackage{leftidx}
\usepackage[text={17cm,24.5cm},centering]{geometry}
\usepackage{textcomp}
\usepackage{wasysym}

\numberwithin{equation}{section}   

\def \be {\begin{equation}}
\def \ee {\end{equation}}
\def \ba {\begin{array}}
\def \ea {\end{array}}
\def \bea{\begin{eqnarray}}
\def \eea{\end{eqnarray}}

\def \a {\alpha}
\def \b {\beta}
\def \g {\gamma}

\def \d {\delta}
\def \D {\Delta}
\def \dg {\dagger}
\def \e {\epsilon}

\def \m {\mu}

\def \l {\lambda}
\def \L {\Lambda}

\def \s {\sigma}

\def \o {\omega}
\def \O {\Omega}

\def \Th {\Theta}
\def \t {\tau}

\def \mC {\mathcal C}

\def \mE {\mathcal E}
\def \mF {\mathcal F}
\def \mG {\mathcal G}

\def \mI {\mathcal I}

\def \mO {\mathcal O}

\def \mQ {\mathcal Q}
\def \mR {\mathcal R}
\def \mS {\mathcal S}
\def \mT {\mathcal T}

\def \p {\partial}

\def \lt {\left}
\def \rt {\right}

\def \td {\tilde}

\def \inf {\infty}

\def \Re {{\textrm{Re}}}

\def \Tr {{\textrm{Tr}}}

\def \diag {{\textrm{diag}}}

\def \and {{\textrm{and}}}

\begin{document}
\begin{titlepage}
	
	\title{\textbf {Charged R\'enyi negativity of massless free bosons}}
	\author{Hui-Huang Chen\footnote{chenhh@jxnu.edu.cn}~,}
	\date{}
	
	\maketitle
	\underline{}
	\vspace{-12mm}
	
	\begin{center}
		{\it
             College of Physics and Communication Electronics, Jiangxi Normal University,\\ Nanchang 330022, China\\
		}
		\vspace{10mm}
	\end{center}
	\begin{abstract}
	 In this paper, we consider the computation of charged moments of the reduced density matrix of two disjoint intervals in the $1+1$ dimensional free compactified boson conformal field theory (CFT) by studying the four-point function of the fluxed twist fields. We obtained the exact scaling function of this four-point function and discussed its decompactification limit. This scaling function was used to obtain the charged moments of the partial transpose which we refer as charged R\'enyi negativity.  These charged moments and the charged moments of the partial transpose are essential for the problem of symmetry decomposition of the corresponding entanglement measures. We test our analytic formula against exact numerical computation in the complex harmonic chain, finding perfect agreements.
	\end{abstract}
	
\end{titlepage}

\thispagestyle{empty}

\newpage

\tableofcontents
\section{Introduction}
In recent years, concepts and ideas coming from quantum information theory have played crucial roles in connecting condensed matter theory and high energy physics \cite{Amico:2007ag, Calabrese:2009qy, Nishioka:2009un}. In quantum many-body systems, one of the most important results on entanglement is the so-called area law (see \cite{Eisert:2008ur} for a review). In high energy physics, Ryu-Takayanagi formula \cite{Ryu:2006bv} firstly recovers the amazing relationship between entanglement and spacetime. People believe that ideas from quantum information theory would help us solve the information paradox of black holes \cite{Hawking:1974sw} ultimately (see \cite{Almheiri:2020cfm} as a review for recent progress). Among all these progress, the entanglement entropies are the most successful entanglement measures to characterize the bipartite entanglement of a pure state. Given a system in a pure state $\ket{\Psi}$, and taking the bipartition of the system $A$ and its complement $B$, the Hilbert space of the full system factorizes as $\mathcal{H}=\mathcal{H}_A\otimes\mathcal{H}_B$. The reduced density matrix (RDM) of subsystem $A$ is defined as $\rho_A=\Tr_B\ket{\Psi}\bra{\Psi}$. From the moments of $\rho_A$, i.e. $\Tr\rho_A^n$, one can obtain the Von Neumann entropy through the replica trick \cite{Calabrese:2004eu}
\be
S_A\equiv-\Tr(\rho_A\log\rho_A)=-\frac{\p}{\p n}\Tr\rho_A^n\Big|_{n=1}
\ee
\par
\begin{figure}
        \centering
        {\includegraphics[width=15cm]{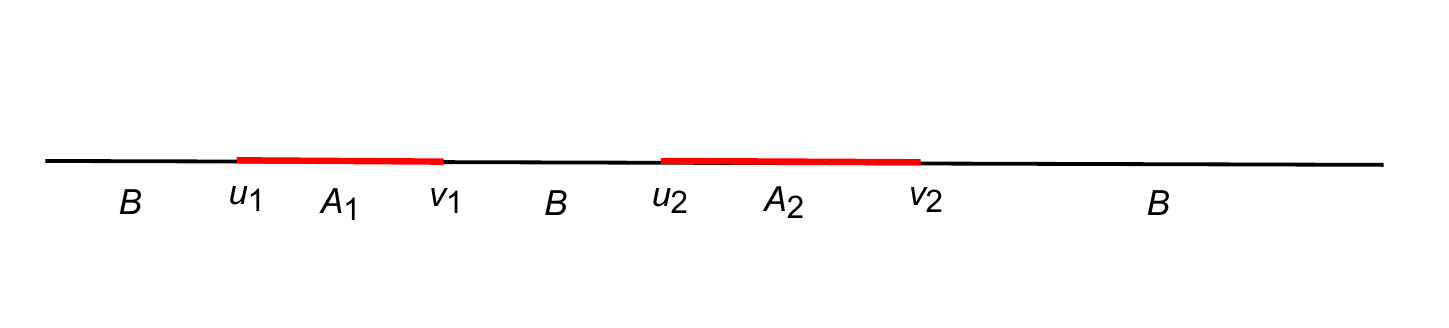}}
        \caption{In 1+1 dimensional spacetime, the time slice is one dimensional, we will consider the density matrix for the field theory on the subsystem $A=A_1\cup A_2$ upon tracing out the local degree of freedom in environment $B$. The endpoints of sub subsystem $A_1,A_2$ is denoted by $u_1,v_1$ and $u_2,v_2$ respectively.}
        \label{fig1}
\end{figure}

 For a mixed state, the entanglement entropies are no longer useful measures of entanglement. Instead, entanglement negativity is one of the most commonly used measures of quantum entanglement between two subsystems $A_1$ and $A_2$ in a generally mixed state \cite{Peres:1996dw, vidal2002computable}. Consider the situation presented in Fig.\ref{fig1}, with the whole system $(A_1\cup A_2)\cup B$ being in a pure state $\ket{\Psi}$. Let $\rho_A$ be the reduced density matrix of subsystem $A=A_1\cup A_2$ obtained by tracing out the ``environment" $B$ as before, and consider the ``partial transpose" of this density matrix, denoted by $\rho_A^{T_2}$, with respect to the Hilbert space corresponding to $A_2$.
Then the entanglement negativity (or negativity for short) is defined as
\be
\mathcal{N}=\frac12(\Tr|\rho_A^{T_2}|-1)=\sum_{\l_i<0}|\l_i|,
\ee
where $\Tr|O|=\Tr\sqrt{O^{\dg}O}$ denotes the trace norm of the operator $O$ and $\l_i$ are the eigenvalues of $\rho_A^{T_2}$.
Logarithmic negativity is another equivalent measure which is defined as
\be
\mE\equiv\log\Tr|\rho_A^{T_2}|.
\ee
To apply to the replica trick on entanglement negativity, it is also useful to introduce the so-called R\'enyi negativity
\be
R_n=\Tr(\rho_A^{T_2})^n.
\ee
\par In quantum field theories (QFT), entanglement negativity has been studied extensively\cite{calabrese2012entanglement, calabrese2013entanglement, kulaxizi2014conformal, bianchini2016branch, blondeau2016universal, Castro-Alvaredo:2019irt}. In recent years, the holographic dual of entanglement negativity has also been proposed and widely investigated\cite{rangamani2014comments, Chaturvedi:2016rft,Chaturvedi:2016rcn,Malvimat:2017yaj, chaturvedi2018holographic, kudler2019entanglement, dong2021holographic}.
\par If our system presents an internal symmetry, the entanglement measures will split into different sectors corresponding to different representations of the symmetry group. After the pioneering work \cite{Goldstein:2017bua}, a lot of progress has been made in this area, including symmetry resolved entanglement entropy \cite{Bonsignori:2019naz, Murciano:2019wdl, Horvath:2020vzs, Fraenkel:2019ykl, Murciano:2020vgh,Bonsignori:2020laa,Capizzi:2021kys}, symmetry resolved relative entropy and trace distance \cite{Chen:2021pls,Capizzi:2021zga}, and symmetry decomposition of entanglement negativity \cite{cornfeld2018imbalance, Murciano:2021djk}, symmetry resolved entanglement entropy in holographic settings \cite{Belin:2013uta, Zhao:2020qmn, Weisenberger:2021eby, Gerbershagen:2021gvc}. All these results are mainly focused on $U(1)$ symmetry. Very recently, symmetry resolved entanglement entropies in Wess-Zumino-Witten (WZW) models were also discussed \cite{Calabrese:2020bys, Milekhin:2021lmq}.
\par In this manuscript, we will first consider the charged moments of the reduced density matrix of two disjoint intervals in 1+1 dimensional free boson conformal field theory (CFT) by the method of fluxed twist fields. As a result, this problem is related to the four-point function of the fluxed twist fields, which is determined by a scaling function. We obtained the exact form of this scaling function and tried to apply this result to studying the problem that how entanglement negativity decomposes under internal symmetry. We finally obtained the charged R\'enyi negativity in this theory and make various numerical checks of our CFT results. The charge imbalance resolved negativity can be obtained by Fourier transforming the charged  R\'enyi negativity and taking the replica limit which has not worked out in this paper due to technical complications.
\par The remaining part of this paper is organized as follows. In section \ref{section2}, we briefly review the path integral representation of negativity in 1+1 dimensional CFT. In section \ref{section3}, we will discuss how entanglement negativity decomposes under internal symmetry. In section \ref{section4}, we will focus on the computation of charged moments of two disjoint intervals and discuss two important regions: the small $x$ region and the decompactification region. In section \ref{section5}, we consider the charged R\'enyi negativity by applying the results obtained in the last section. In section \ref{section6}, we test our analytic formula against exact numerical computation. Finally, we make a conclusion and discussion of our results in section \ref{section7} and some technical details are present in two appendices.
\section{Entanglement negativity in CFT}\label{section2}
\subsection{Moments of reduced density matrix}
Let us briefly review the calculation of R\'enyi entanglement negativity via path integrals in two-dimensional CFTs. We denote $\phi$ as the basic field in the theory. The matrix elements of the density matrix $\rho$ of vacuum are $\rho(\phi,\phi')=\braket{\phi|0}\braket{0|\phi'}$.
Now we consider the subsystem $A$ consists $N$ disjoint interval on the real axis, $A=\cup_{i=1}^N[u_i,v_i]$. The matrix elements of RDM $\rho_A$ are given by tracing out the degree of freedom not in $A$
\be
\rho_A(\phi,\phi')=\frac{1}{Z}\int\mathcal{D}\phi_B\braket{\phi_A=\phi,\phi_B|0}\braket{0|\phi_A=\phi',\phi_B}.
\ee
For integer $n$, the moments of RDM $\rho_A$ are
\be
\Tr\rho_A^n=\frac{1}{Z_1^n}\int\prod_{i=1}^n\mathcal{D}\phi_i\rho_A(\phi_i,\phi_{i+1}),\quad \phi_{n+1}\equiv\phi_1,
\ee
where $Z_1$ is the partition function on the sphere and from the path integral above we should glue the cuts cyclicly along with $A$, which turns out to be the partition function on an $n$-sheeted Riemann surface $\mathcal{R}_{n,N}$ with genus $(n-1)(N-1)$. The simplest and most important example is the entanglement entropy of an interval $A$ of length $l$ in an infinite line. Since the genus of the corresponding Riemann surface for $N=1$ is zero, the entanglement entropy $S_A$ has the universal form
\be
S_A=\frac{c}{3}\log\frac{l}{\e}+\mathrm{const},
\ee
where $c$ is the central charge of the CFT, and $\e$ is the UV cutoff.
\par  For general $N>1$, it's convenient to introduce branch point twist fields in the complex plane to avoid from studying complicated Riemann surfaces. As a result, the partition function on Riemann surface $\mathcal{R}_{n,N}$ can be written as $2N$ point function of branch point twist fields in the complex plane \cite{Calabrese:2004eu}
\be
\Tr\rho_A^n=\langle\prod_{i=1}^N\mathcal{T}_n(u_i)\tilde{\mathcal{T}}_n(v_i)\rangle
\ee
The twist fields $\mathcal{T}_n$ and anti-twist field $\tilde{\mathcal{T}_n}$ are primary operators and they have the same dimension
\be
\D_n=\frac{c}{12}\left(n-\frac{1}{n}\right).
\ee
By conformal map $w(z)=\frac{(u_1-z)(u_N-v_N)}{(u_1-u_N)(z-v_N)}$, the above $2N$ point function only depends on $2N-3$ cross-ratios $x_1=w(v_1),x_2=w(u_2),\cdots,x_{2N-3}=w(v_{N-1})$, then we can write
\be\label{Ninterval}
\Tr\rho_A^n=c_n^N\Big|\frac{\prod_{i<j}(u_j-u_i)(v_j-v_i)}{\prod_{i,j}(v_j-u_i)}\Big|^{2\D_n}\mathcal{F}_{N,n}(x_1,\cdots,x_{2N-3}),
\ee
where $\mathcal{F}_{N,n}(x_1,\cdots,x_{2N-3})$ is sensitive to the full operator content of the theory and usually is very complicated. Only very few results are known \cite{Hubeny:2007re, Calabrese:2009ez, Calabrese:2010he, Tonni:2010pv,Headrick:2010zt,Headrick:2012fk}.  For the free real boson compactified on a circle of radius $\mathrm{R}$, analytic expressions are available for $N=2$ and for integer $n>1$. In this case, there is only one independent cross ration $x=\frac{(u_1-v_1)(u_2-v_2)}{(u_1-u_2)(v_1-v_2)}$. Then eq.~(\ref{Ninterval}) specialised to the case $N=2$ reads
\be\label{Twointerval}
\Tr\rho_A^n=c_n^2\left(\frac{(u_2-u_1)(v_2-v_1)}{(v_1-u_1)(v_2-u_2)(v_2-u_1)(u_2-v_1)}\right)^{2\D_n}\mathcal{F}_{n}(x),
\ee
where $\mF_n(x)\equiv\mF_{2,n}(x)$. The function $\mathcal{F}_{n}(x)$ is \cite{Calabrese:2009ez}(parametrized in terms of $\eta=\mathrm{R}^2/2$)
\be\label{F2n}
\mathcal{F}_n(x)=\frac{\Theta(\bold{0}|\eta\tau)\Theta(\bold{0}|\tau/\eta)}{\Theta(\bold{0}|\tau)^2},
\ee
where $\tau$ is $(n-1)\times(n-1)$ matrix with the following entries:
\be\label{tau}
\tau_{rs}=\frac{2\mathrm{i}}{n}\sum_{k=1}^{n-1}\sin(\frac{\pi k}{n})\frac{_2F_1(k/n,1-k/n;1;1-x)}{_2F_1(k/n,1-k/n;1;x)}\cos[\frac{2\pi k}{n}(r-s)].
\ee
While $\Theta$ is the Siegel theta function defined by
\be
\Theta(\mathbf{z}|M)=\sum_{\mathbf{m}\in \mathbb{Z}^{n-1}}e^{\mathrm{i}\pi\mathbf{m}^t\cdot M\cdot\mathbf{m}+2\pi \mathrm{i}\mathbf{m}^t\cdot\mathbf{z}},
\ee
which is a function of the $(n-1)$ dimensional vector $\mathbf{z}$ and of the $(n-1)\times(n-1)$ matrix $M$ which must be symmetric and with positive imaginary part.
In the remaining part of this paper, we will mainly focus on the case $N=2$.
\subsection{The repica trick for negativity}
In this section, we briefly review the computation of entanglement negativity in 1+1 dimensional CFT using twist fields. Let $\ket{e_i^{(1)}}$ and $\ket{e_j^{(2)}}$ be two arbitrary bases of the Hilbert spaces associated to the degree of freedom on $A_1$ and $A_2$ respectively. The partial transpose (with respect to the second space) of $\rho_A$ is defined as
\be
\bra{e_i^{(1)}e_j^{(2)}}\rho_A^{T_2}\ket{e_k^{(1)}e_l^{(2)}}=\bra{e_i^{(1)}e_l^{(2)}}\rho_A\ket{e_k^{(1)}e_j^{(2)}}.
\ee
The R\'enyi negativity is defined as
\be
R_n=\Tr\{(\rho_A^{T_2})^n\}=\sum_{\l_i}\l_i^n,
\ee
and entanglement negativity is obtained by taking the limit
\be
\mathcal{N}=\lim_{n\rightarrow\frac12}\frac12(R_{2n}-1).
\ee
\par As mentioned in the last subsection, $\Tr\rho_A^n$ for the union of two disjoint interval $A=[u_1,v_1]\cup[u_2,v_2]$ is given by the correlator $\langle\mathcal{T}_n(u_1)\tilde{\mathcal{T}}_n(v_1)\mathcal{T}_n(u_2)\tilde{\mathcal{T}}_n(v_2)\rangle$.Taking partial transpose with respect to the interval $A_2$ means the endpoints of $A_2$ are exchanged while $A_1$ stay untouched. According to standard procedure, the path integral representation of $\Tr\{(\rho_A^{T_2})^n\}$ is proportional to the partition function on the Riemann surface $\tilde{\mathcal{R}}_{n,2}$. $\tilde{\mathcal{R}}_{n,2}$ differs from $\mathcal{R}_{n,2}$ only by the exchange $u_2\leftrightarrow v_2$, and this immediately translates in terms of correlation function of twist fields as $\Tr(\rho_A^{T_2})^n=\langle\mathcal{T}_n(u_1)\tilde{\mathcal{T}}_n(v_1)\tilde{\mathcal{T}}_n(u_2)\mathcal{T}_n(v_2)\rangle$. Since the exchange $u_2\leftrightarrow v_2$ sends $x\rightarrow x/(x-1)$, the period matrix $\tilde{\tau}$ of $\tilde{\mathcal{R}}_{n,2}$ for $x\in (0,1)$ is given by
\be
\tilde{\tau}(x)=\tau(x/(x-1))=\mathcal{R}+i\mathcal{I},
\ee
where $\tau$ is the period matrix of $\mathcal{R}_{n,2}$ whose elements have been given in eq.~(\ref{tau}) and we denote the real and imaginary part of $\tilde{\tau}(x)$ by $\mathcal{R}$ and $\mathcal{I}$ respectively.
\par Since the moments of the partial transposed RDM can be written as four-point functions of twist fields. This four-point function still have the same form with equation eq.~(\ref{Twointerval})
\be
\Tr(\rho_A^{T_2})^n=c_n^2\left(\frac{(u_2-u_1)(v_2-v_1)}{(v_1-u_1)(v_2-u_2)(v_2-u_1)(u_2-v_1)}\right)^{2\D_n}\mathcal{G}_{n}(x),
\ee
and we have the following relation
\be
\mathcal{G}_{n}(x)=(1-x)^{4\D_n}\mathcal{F}_{n}\left(\frac{x}{x-1}\right).
\ee
The result eq.~(\ref{F2n}) is valid only for $0<x<1$. For generic $x\in\mathbb{C}$, it can be written as
\be\label{F2nGeneric}
\mathcal{F}_n(x,\bar{x})=\frac{\Theta(\mathbf{0}|T)}{|\Theta(\mathbf{0}|\tau)|^2},
\ee
where the $2(n-1)\times 2(n-1)$ symmetric matrix $T$ is
\be
T=\begin{pmatrix}
i\eta\mI&\mR\\
\mR&i\mI/\eta
\end{pmatrix},
\ee
When $x\in(0,1)$, the period matrix $\tau(x)$ is purely imaginary. Then $\mR=0$ and the Siegel theta function in eq.~(\ref{F2nGeneric}) factorizes, giving back the result eq.~(\ref{F2n}).
\section{Symmetry decomposition of negativity}\label{section3}
Let us first review some basic facts about the symmetry resolution of the entanglement entropy.
In this section and the remaining part of this paper, we will assume our system presents a global $U(1)$ symmetry $Q$ and assuming that $[\rho, Q]=0$, which implies $[\rho_A,Q_A]=0$. We take the subsystem $A$ consist of two disjoint region $A=A_1\cup A_2$. Then $\rho_A$ admits charge decomposition according to eigenvalues $q$ of local charge $Q_A$
\be
\rho_A=\oplus_q\Pi_q\rho_A=\oplus_qp(q)\rho_A(q),\quad p(q)=\Tr(\Pi_q\rho_A).
\ee
The symmetry resolved R\'enyi entropies are then defined as
\be
S_n(q)=\frac{1}{1-n}\log\Tr[\rho_A(q)]^n.
\ee
\par It's very difficult to compute the symmetry resolved entropy using the above definition due to the non-local feature of the projector $\Pi_q$. However, we can bypass this difficulty by using the Fourier representation of the projector and focusing on the charged moments of $\rho_A$,
\be\label{Znalpha}
Z_n(\m)=\Tr(e^{i\m Q_A}\rho_A^n).
\ee
Then it's sufficient to compute the Fourier transform
\be
Z_n(q)=\int_0^{2\pi}\frac{d\m}{2\pi}e^{-iq\m}Z_n(\m)
\ee
to obtain the entropies of the sector of charge $q$ as
\be
S_n(q)=\frac{1}{1-n}\log\left[\frac{Z_n(q)}{Z_1(q)^n}\right].
\ee
\par Now we turn to the symmetry decomposition of entanglement negativity under $U(1)$ charge $Q$. Since the charge $Q$ is local, we can write $Q_A=Q_1+Q_2$. From the relation $[\rho_A,Q_A]=0$, performing a partial transposition with respect to the second region $A_2$ of subsystem $A$, we obtain
\be
[\rho_A^{T_2},\mQ]=0,\quad \mQ\equiv Q_1-Q_2^{T_2},
\ee
where we have introduced the charge imbalance operator $\mQ$ and we will denote it's eigenvalues as $\mathrm{q}$ to make a distinction with the eigenvalues of $Q_A$.
Then $\rho_A^{T_2}$ has a block matrix form, each block was characterized by different eigenvalues $\mathrm{q}$ of the imbalance operator $\mQ$
\be
\rho_A^{T_2}=\oplus_{\mathrm{q}}\Pi_{\mathrm{q}}\rho_A^{T_2}.
\ee
we define the normalized charge imbalance partially transposed density matrix
\be
\rho_A^{T_2}({\mathrm{q}})=\frac{\Pi_{\mathrm{q}}\rho_A^{T_2}}{\Tr(\Pi_{\mathrm{q}}\rho_A^{T_2})}.
\ee
Then we can write
\be
\rho_A^{T_2}({\mathrm{q}})=\oplus_{\mathrm{q}} p({\mathrm{q}})\rho_A^{T_2}({\mathrm{q}}),
\ee
where $p({\mathrm{q}})=\Tr(\Pi_{\mathrm{q}}\rho_A^{T_2})$ is the probability of finding ${\mathrm{q}}$ as the outcome of measurement of $\mQ$.
\par The charge imbalance resolved negativity is defined as
\be
\mathcal{N}({\mathrm{q}})=\frac{1}{2}(\Tr|\rho_A^{T_2}({\mathrm{q}})|-1).
\ee
The total negativity is given by the sum of charge imbalance resolved negativity weighted by the corresponding probability
\be
\mathcal{N}=\sum_{\mathrm{q}}p({\mathrm{q}})\mathcal{N}({\mathrm{q}}).
\ee
The charge imbalance resolved R\'enyi negativity is defined as
\be
R_n({\mathrm{q}})=\Tr[(\rho_A^{T_2}({\mathrm{q}}))^n]=\frac{1}{p({\mathrm{q}})^n}\Tr[\Pi_{\mathrm{q}}(\rho_A^{T_2})^n].
\ee
Then the charge imbalance entanglement negativity can be obtained by taking the limit
\be
\mathcal{N}({\mathrm{q}})=\lim_{n\rightarrow\frac12}\frac12(R_{2n}({\mathrm{q}})-1).
\ee
The projection operator $\Pi_\mathrm{q}$ has the following integral representation
\be
\Pi_{\mathrm{q}}=\int_{0}^{2\pi}\frac{d\m}{2\pi}e^{-i\m {\mathrm{q}}}e^{i\m(Q_1-Q_2^{T_2})}.
\ee
We can first compute the charged  R\'enyi negativity
\be\label{Rnmu}
R_n(\m)=\Tr[(\rho_A^{T_2})^ne^{i\m(Q_1-Q_2^{T_2})}],
\ee
then the charge imbalance resolved R\'enyi negativity are related by Fourier transform
\be
R_n({\mathrm{q}})=\frac{\int_0^{2\pi}\frac{d\m}{2\pi} e^{-i\m {\mathrm{q}}}R_n(\m)}{[\int_0^{2\pi}\frac{d\m}{2\pi} e^{-i\m {\mathrm{q}}}R_1(\m)]^n}.
\ee
\section{Charged moments of two disjoint interval in free boson CFT}\label{section4}
\subsection{Fluxed twist fields and charged moments}
In a generic two-dimensional QFT, according to the path integral representation of the charged moments $Z_n(\m)$ defined in eq.~(\ref{Znalpha}), we can treat it as the partition function on the Riemann surface $\mR_{n,N}$ pierced by an Aharonov-Bohm flux, such that the total phase accumulated by the field upon going through the entire surface is $\m$. We denote this fluxed Riemann surface by $\mR_{n,N}^{(\m)}$. The presence of the flux corresponds to impose twist boundary condition.
This boundary condition fuses with twist field at the endpoints of subsystem $A$ can be implemented by two local fields $\mT_{n,\m}$ and $\tilde{\mT}_{n,\m}$ termed as fluxed twist fields and fluxed anti-twist fields. These fluxed twist fields take into account not only the internal permutational symmetry among the replicas but also the presence of the flux. The partition function on $\mR_{n,N}^{(\m)}$ is related to the $2N$-point function of the fluxed twist operators.
\par Studying field theory on the complicated fluxed Riemann surface usually turns out to be very hard. It's more convenient to move the topology of the world-sheet to the target space where the fields lie. In other words, we can work on a theory defined on a single complex plane formed by $n$ independent copies of the original theory. In this multi-copy theory, we have to deal with fields with $n$-component and we denote $\phi_j$ as the field on the $j$-th copy.
For subsystem $A$ made of two disjoint intervals, we have
\be
Z_n(\m)=\langle\mathcal{T}_{n,\m}(u_1)\tilde{\mathcal{T}}_{n,\m}(v_1)\mathcal{T}_{n,\m}(u_2)\tilde{\mathcal{T}}_{n,\m}(v_2)\rangle.
\ee
\par Upon crossing the cut $A$, the field $\phi_i$ transforms as $\phi_i\rightarrow (T_{\m})_{ij}\phi_j$, where the matrix elements of the transformation matrix $T_{\m}$ are given by $(T_{\m})_{ij}=e^{\mathrm{i}\m/n}\d_{i+1,j}$ with boundary condition $n+1\equiv 1$. The eigenvectors of the matrix $T_{\m}$ are
\be
\varphi_k=\sum_{j=1}^ne^{2\pi i\frac{k}{n}j}\phi_j,\qquad k=0,1,2,\cdots,n-1,
\ee
with corresponding eigenvalues given by $\l_k=e^{\mathrm{i}\frac{\m}{n}+2\pi\mathrm{i}\frac{k}{n}}$. In other words, in the space span by the fields $\varphi_k$, the action of the fluxed twist field decoupled, i.e. they act on each $k$ independently. Thus we can write
\be
\mT_{n,\m}=\prod_{k=0}^{n-1}\mT_{n,\m,k},\qquad \td{\mT}_{n,\m}=\prod_{k=0}^{n-1}\td{\mT}_{n,\m,k}.
\ee
Thus, the partition function $Z_n(\m)$ also factorize as
\be
Z_n(\m)=\prod_{k=0}^{n-1}Z_{n,k}(\m)=\prod_{k=0}^{n-1}\langle\mathcal{T}_{n,\m,k}(u_1)\tilde{\mathcal{T}}_{n,\m,k}(v_1)\mathcal{T}_{n,\m,k}(u_2)\tilde{\mathcal{T}}_{n,\m,k}(v_2)\rangle.
\ee
\par Now let us consider a concrete model. Namely, we will consider the complex bosonic free compactified field with Euclidean action
\be\label{action}
\mS[\phi]=\frac{g}{4\pi}\int dzd\bar{z}\p_z\phi\p_{\bar{z}}\phi.
\ee
In the multi-copy theory, we have the $n$-component field $\Phi=(\phi_1,\phi_2,\cdots,\phi_n)^T$, each component $\phi_j$ are free but compactified on a circle. Since the field is complex, the target space is a torus with radius $\mathrm{R}_1$ and $\mathrm{R}_2$. Taking the non-trivial winding into account, encircling the endpoints of $A$ will lead to
\be\label{Monodromy}
\begin{split}
\phi_j(e^{2\pi\mathrm{i}}z_i,e^{-2\pi\mathrm{i}}\bar{z}_i)=e^{\mathrm{i}\m/n}\phi_{j-1}(z_i,\bar{z}_i)+m_{j,1}\mathrm{R}_1+\mathrm{i} m_{j,2}\mathrm{R}_2,\qquad m_{j,1},m_{j,2}\in \mathbb{Z}\\
\phi_j(e^{2\pi \mathrm{i}}\td{z}_i,e^{-2\pi\mathrm{i}}\bar{\td{z}}_i)=e^{-\mathrm{i}\m/n}\phi_{j+1}(\td{z}_i,\bar{\td{z}}_i)+\td{m}_{j,1}\mathrm{R}_1+\mathrm{i} \td{m}_{j,2}\mathrm{R}_2,\qquad \td{m}_{j,1},\td{m}_{j,2}\in \mathbb{Z},
\end{split}
\ee
where $z_i\equiv z-u_i,\td{z}_i\equiv z-v_i$ for $i=1,2$. In this theory, $\mT_{n,\m,k}$ and $\td{\mT}_{n,\m,k}$ are bosonic $U(1)$ twist field with dimension (see appendix A for details)
\be
\D_{n,k,\m}=\lt(\frac{k}{n}+\frac{\m}{2\pi n}\rt)\lt(1-\frac{k}{n}-\frac{\m}{2\pi n}\rt).
\ee
Then, the dimension of the fluxed twist fields $\mT_{n,\m}$ can be obtained by
\be
\D_{n,\m}=\sum_{k=0}^{n-1}\D_{n,k,\m}=\frac{1}{6}\lt(n-\frac{1}{n}\rt)-\frac{\m^2}{4\pi^2n}+\frac{\m}{2\pi n}.
\ee
\par We can simplify this non-trivial monodromy a bit by introducing $\varphi_k=\sum_{j=1}^ne^{2\pi\mathrm{i}jk/n}\phi_j$. Then eq.~(\ref{Monodromy}) becomes
\be
\begin{split}
\varphi_k(e^{2\pi\mathrm{i}}z_i,e^{-2\pi\mathrm{i}}\bar{z}_i)=e^{\mathrm{i}\m/n+2\pi \mathrm{i}k/n}\varphi_{k}(z_i,\bar{z}_i)+\mathrm{R}\sum_{j=1}^ne^{2\pi\mathrm{i}jk/n}m_j,\\
\varphi_k(e^{2\pi\mathrm{i}}\td{z}_i,e^{-2\pi\mathrm{i}}\bar{\td{z}}_i)=e^{-\mathrm{i}\m/n-2\pi \mathrm{i}k/n}\varphi_{k}(\td{z}_i,\bar{\td{z}}_i)+\mathrm{R}\sum_{j=1}^ne^{2\pi\mathrm{i}jk/n}\td{m}_j.
\end{split}
\ee
Here, for simplicity we set $\mathrm{R}_1=\mathrm{R}_2=\mathrm{R}$ and $m_j,\td{m}_j\in\mathbb{Z}+\mathrm{i}\mathbb{Z}$. Consequently, the target space of the field $\varphi_k$ is a complicated two dimensional lattice $\L$
\be
\L=\lt\{\mathrm{R}\sum_{j=1}^ne^{2\pi\mathrm{i}jk/n}m_j\Big|m_j\in\mathbb{Z}+\mathrm{i}\mathbb{Z}\rt\}.
\ee
\par In order to calculate the partition function, we can split $\varphi_k$ into a classical and a quantum part $\varphi_k=\varphi^{\mathrm{cl}}_k+\varphi^{\mathrm{qu}}_k$. The classical part takes into account the nontrivial structure of the target space
\be
\varphi^{\mathrm{cl}}_k(e^{2\pi\mathrm{i}}z_i,e^{-2\pi\mathrm{i}}\bar{z}_i)=e^{2\pi\mathrm{i}a}\varphi^{\mathrm{cl}}_k(z_i,\bar z_i)+\l,
\ee
where $\l\in\L$, while the quantum part is transparent to it
\be
\varphi^{\mathrm{qu}}_k(e^{2\pi\mathrm{i}}z_i,e^{-2\pi\mathrm{i}}\bar{z}_i)=e^{2\pi\mathrm{i}a}\varphi^{\mathrm{qu}}_k(z_i,\bar z_i),
\ee
where we have defined $a\equiv\frac{k}{n}+\frac{\m}{2\pi n}$ and similar relation holds when encircling around $\td{z}_i$.
The computation of quantum part is already done in \cite{dixon1987conformal} and we report its derivation in the appendix \ref{AppendixA}
\be
Z^{\mathrm{qu}}_{n,k}(\m)=\mathrm{const}\left(\frac{(u_2-u_1)(v_2-v_1)}{(v_1-u_1)(v_2-u_2)(v_2-u_1)(u_2-v_1)}\right)^{2\D_{n,k,\m}}\frac{1}{I_{n,k,\m}(x)},
\ee
where
\be
I_{n,k,\m}(x)=2F_a(x)F_{a}(1-x)=2\beta_a[F_a(x)]^2,
\ee
and
\be
F_y(x)\equiv{_2F_1(y,1-y;1;x)},\quad \b_y(x)\equiv\frac{F_y(1-x)}{F_y(x)}.
\ee
The action for a given classical configuration is given by (eq.~(\ref{ClassicalAction}) with $\a_a=0$)
\be
\mS^{\mathrm{cl}}_a=\frac{2g\pi\sin\pi a}{n\beta_a}[|\xi_1|^2+\beta_a^2|\xi_2|^2],
\ee
where $\xi_{1},\xi_2\in \L$.
Then the partition function can be written as
\be
Z_n(\m)=\sum_{m\in\mathbb{Z}^{2n}}\prod_{k=0}^{n-1}Z_a^{\mathrm{qu}}Z_a^{\mathrm{cl}}.
\ee
The classical part of the partition function reads
\be\label{Zcl0}
Z^{\mathrm{cl}}_n(\m)=\sum_{m\in\mathbb{Z}^{2n}}\prod_{k=0}^{n-1}Z_a^{\mathrm{cl}}=\sum_{\mathbf{m}\in\mathbb{Z}^{2n}}\prod_{k=0}^{n-1}\exp\Big\{-\frac{2g\pi\sin\pi a}{n}[|\xi_1|^2\b_a+\b_a^{-1}|\xi_2|^2]\Big\}.
\ee
Since the quantum part does not depend on $\mathbf{m}$, the function $\mF_n(\m,x)$ can be written as
\be\label{Fnmux0}
\mF_n(\m,x)=\prod_{k=0}^{n-1}\frac{\mathrm{const}}{\b_a[F_a(x)]^2}\sum_{\mathbf{m}\in\mathbb{Z}^{2n}}\prod_{k=0}^{n-1}\exp\Big\{-\frac{2g\pi\sin\pi a}{n}[|\xi_1|^2\b_a+\b_a^{-1}|\xi_2|^2]\Big\}.
\ee
Given $\xi_i=\mathrm{R}\sum_{r=1}^ne^{2\pi\mathrm{i}kr/n}(m_{r,1}^{(i)}+\mathrm{i}m_{r,2}^{(i)}),i=1,2$,
we have
\be\label{xi}
|\xi_i|^2=\mathrm{R}^2\sum_{r,s=1}^n\Big[\sum_{j=1}^2m^{(i)}_{r,j}m^{(i)}_{s,j}\Big[\cos\frac{2\pi(r-s)k}{n}\Big]
+(m^{(i)}_{r,1}m^{(i)}_{s,2}-m^{(i)}_{s,1}m^{(i)}_{r,2})\sin\Big[\frac{2\pi(r-s)k}{n}\Big]\Big].
\ee
Then substitute eq.~(\ref{xi}) into eq.~(\ref{Zcl0}) and after some simplification, one can rewrite the result in terms of Siegel theta functions. Since the computation is almost the same as \cite{Calabrese:2009ez}, we just report the final result here
\be
Z^{\mathrm{cl}}_n(\m)=[\Theta(\mathbf{0}|\eta\O)\Theta(\mathbf{0}|\eta\td{\O})]^2,
\ee
where $\O$ and $\tilde{\O}$ are $n\times n$ matrix with elements
\be
\O_{rs}=\frac{2\mathrm{i}}{n}\sum_{k=0}^{n-1}\sin(\pi a)\beta_a\cos[\frac{2\pi k}{n}(r-s)],
\ee
\be
\tilde{\O}_{rs}=\frac{2\mathrm{i}}{n}\sum_{k=0}^{n-1}\sin(\pi a)\frac{1}{\beta_a}\cos[\frac{2\pi k}{n}(r-s)],
\ee
and $\eta\equiv g\mathrm{R}^2$.
The spectrum of the matrices $\O$ and $\td{\O}$ are given by
\be
\{\o_p|p=1,2,\cdots,n-1\}\cup\Big\{\o_n\equiv\mathrm{i}\sin\lt(\frac{\m}{2n}\rt)\b_{\frac{\m}{2\pi n}}\Big\}
\ee
and
\be
\{\tilde{\o}_p|p=1,2,\cdots,n-1\}\cup\Big\{\td{\o}_n\equiv\mathrm{i}\sin\lt(\frac{\m}{2n}\rt)\Big/\b_{\frac{\m}{2\pi n}}\Big\}
\ee
respectively, where for $p=1,2,\cdots,n-1$, we have
\be
\o_p=\mathrm{i}\sin\left[\pi\left(\frac{p}{n}-\frac{\m}{2\pi n}\right)\right]\b_{\frac{p}{n}-\frac{\m}{2\pi n}}+\mathrm{i}\sin\left[\pi\left(\frac{p}{n}+\frac{\m}{2\pi n}\right)\right]\b_{\frac{p}{n}+\frac{\m}{2\pi n}},
\ee
\be
\tilde{\o}_p=\mathrm{i}\sin\left[\pi\left(\frac{p}{n}-\frac{\m}{2\pi n}\right)\right]\Big/\b_{\frac{p}{n}-\frac{\m}{2\pi n}}+\mathrm{i}\sin\left[\pi\left(\frac{p}{n}+\frac{\m}{2\pi n}\right)\right]\Big/\b_{\frac{p}{n}+\frac{\m}{2\pi n}}.
\ee
The matrices $\O$ and $\tilde{\O}$ have a common eigenbasis whose normalized eigenvectors can be written as
\be
(e_p)_q=\frac{1}{\sqrt{n}}e^{2\pi\mathrm{i}pq/n},\quad p,q=1,2,\cdots,n.
\ee
Then $\O$ and $\td{\O}$ can be diagonalized simultaneously by matrix $U$ whose elements are given by $U_{pq}=(e_p)_q$.
\par We finally have
\be\label{Fnmux}
\mathcal{F}_n(\m,x)=s_{n,\m}\frac{[\Theta(\mathbf{0}|\eta\O)\Theta(\mathbf{0}|\eta\tilde{\O})]^2}{\prod_{k=0}^{n-1}\beta_a[F_a(x)]^2}.
\ee
Here $s_{n,\m}$ can be fixed by requiring $\mathcal{F}_n(\m,0)=1$ as (See appendix \ref{AppendixB} for details)
\be
s_{n,\m}=2n\eta^n\cos^{n-1}\left(\frac{\m}{2n}\right)\sin\left(\frac{\m}{2n}\right).
\ee
Using identities of the $\Theta$ function in the appendix \ref{AppendixB}, we find another expression of $\mF_{n}(\m,x)$
\be
\mathcal{F}_n(\m,x)=s_{n,\m}\frac{[\Theta(\mathbf{0}|\eta\O)\Theta(\mathbf{0}|\hat{\O}/\eta)]^2}{\prod_{p=1}^n(-\mathrm{i}\eta\td{\o}_p)\prod_{k=0}^{n-1}\beta_a[F_a(x)]^2},
\ee
where
\be
\hat{\O}_{rs}=-\td{\O}^{-1}_{rs}=\frac{-1}{n}\sum_{p=1}^n\td{\o}_p^{-1}\cos[\frac{2\pi p}{n}(r-s)].
\ee
\subsection{Small $x$ regime}
For $x\rightarrow0$, the function $\b_a(x)$ behave as
\be
\b_a(x)=-\frac{\sin\pi a}{\pi}(\log x+f_a)+\mO(x),\quad f_a\equiv 2\g_E+\psi(a)+\psi(1-a),
\ee
where $\g_E$ is the Euler constant and $\psi(z)$ is the diagamma function. Then
\be
\Th(0|\eta\O)=1+\sum_{\mathbf{m}\in\mathbb{Z}^n/{\mathbf{0}}}x^{\frac{2\eta}{n}\sum_{k=0}^{n-1}\sin^2(\pi a)\mathbf{m}^{\mathrm{t}}\cdot C_{k/n}\cdot\mathbf{m}}e^{\frac{2\eta}{n}\sum_{k=0}^{n-1}\sin^2(\pi a)f_a\mathbf{m}^{\mathrm{t}}\cdot C_{k/n}\cdot\mathbf{m}}+\cdots,
\ee
where $(C_{k/n})_{rs}\equiv \cos[\frac{2\pi k}{n}(r-s)]$
and the following relation holds
\be\label{power}
\frac{2}{n}\sum_{k=0}^{n-1}\sin^2(\pi a)\mathbf{m}^{\mathrm{t}}\cdot C_{k/n}\cdot\mathbf{m}=\sum_{j=1}^nm_j^2-\cos\lt(\frac{\m}{n\pi}\rt)\sum_{j=1}^nm_jm_{j+1},
\ee
where $m_{n+1}\equiv m_1$. This expression is obviously minimized when the non-zero vector $\mathbf{m}$ has only one entry equals to $\pm1$ and all other elements are zero. There are $2n$ such vectors for which eq.~(\ref{power}) equals to $1$. All these $2n$ vectors gives the same value for $\mathbf{m}^{\mathrm{t}}\cdot C_{k/n}\cdot\mathbf{m}$, and the result is 1.
Using the following integral representation of diagamma function
\be
\psi(z)+\g_E=\int_0^{\inf}dt\frac{e^{-t}-e^{-zt}}{1-e^{-t}},
\ee
we find the formula
\be
\begin{split}
\frac{2}{n}\sum_{k=0}^{n-1}\sin^{2}(\pi a)f_a=f_{\frac{\m}{2\pi}}-2\log n
+\Re~B_{\theta}\lt(\frac{\m}{2\pi},0\rt)
+\Re~B_{\theta}\lt(1-\frac{\m}{2\pi},0\rt),
\end{split}
\ee
where $B_z(a,b)$ is the incomplete Beta function
\be
B_z(a,b)=\int_0^{z}t^{a-1}(1-t)^{b-1}dt,
\ee
and $\theta\equiv e^{2\pi\mathrm{i}/n}$.
\par Then the coefficient of $x^{\eta}$ is $2n^{1-2\eta}\exp\{\eta[f_{\m/2\pi}
+\Re~B_{\theta}(\m/2\pi,0)
+\Re~B_{\theta}(1-\m/2\pi,0)]\}$.
\par To find the small $x$ expansion of $\Theta(\mathbf{0}|\hat{\O}/\eta)$, we first compute
\be
\td{\o}_p^{-1}=
\begin{cases}
\frac{\mathrm{i}}{2\pi}\lt(\log x+\frac12\lt(f_{a^{+}}+f_{a^{-}}\rt)\rt)+\mO\lt(\frac{1}{\log x}\rt),& p=1,2,\cdots,n-1\\
\frac{\mathrm{i}}{2\pi}\lt(\log x+f_{\frac{\m}{2\pi n}}\rt)+\mO\lt(\frac{1}{\log x}\rt),&p=n
\end{cases}
\ee
where $a^{\pm}\equiv \frac{k}{n}\pm\frac{\m}{2\pi n}$.
Then
\be
\Th(0|\hat{\O}/\eta)=1+\sum_{\mathbf{m}\in\mathbb{Z}^n/{\mathbf{0}}}x^{\frac{1}{2n\eta}\sum_{k=1}^{n}\mathbf{m}^{\mathrm{t}}\cdot C_{k/n}\cdot\mathbf{m}}e^{\frac{1}{4n\eta}\sum_{k=1}^{n-1}(f_{a^{+}}+f_{a^{-}})\mathbf{m}^{\mathrm{t}}\cdot C_{k/n}\cdot\mathbf{m}}
e^{\frac{1}{2n\eta}f_{\frac{\m}{2\pi n}}\mathbf{m}^{\mathrm{t}}\cdot C_{n/n}\cdot\mathbf{m}}+\cdots
\ee
From the relation
\be\label{power1}
\frac{1}{2n}\sum_{k=1}^{n}\mathbf{m}^{\mathrm{t}}\cdot C_{k/n}\cdot\mathbf{m}=\frac12\sum_{j=1}^nm_j^2,
\ee
it's easy to see that this expression is minimized when the non-zero vector $\mathbf{m}$ has only one entry equals to $\pm1$ and all other elements are zero. There are $2n$ such vectors for which eq.~(\ref{power1}) equals to $\frac12$.
From the relation
\be
\frac{1}{2}\sum_{k=1}^{n-1}(f_{a^+}+f_{a^-})+f_{\frac{\m}{2\pi n}}=n(f_{\frac{\m}{2\pi}}-2\log n),
\ee
we see that the coefficient of $x^{\frac{1}{2\eta}}$ is $2n\exp[\frac{1}{2\eta}(f_{\frac{\m}{2\pi}}-2\log n)]$.
Therefore we get the small $x$ behavior of the scaling function
\be
\mF_{n}(\m,x)=1+\xi_{n,\m}x^{\a}+\cdots
\ee
where
\be
    \xi_{n,\m}=
\begin{cases}
    4n\exp[\frac{1}{2\eta}(f_{\frac{\m}{2\pi}}-2\log n)],& \text{if } \frac{1}{2\eta}\leq\eta \\
    4n^{1-\eta}\exp\{\eta[f_{\frac{\m}{2\pi}}+\Re~B_{\theta}\lt(\frac{\m}{2\pi},0\rt)+\Re~B_{\theta}\lt(1-\frac{\m}{2\pi},0\rt)]\},              & \text{otherwise}
\end{cases}
\ee
and $\a=\min(\eta,\frac{1}{2\eta})$.
\subsection{Decompactification regime}
For fixed $x$, in the limit of large $\eta$, we have $\Theta(\mathbf{0}|\eta\O)=1+\cdots$ and $\Theta(\mathbf{0}|\eta\td{\O})=1+\cdots$, where $\cdots$ denotes terms which vanish when taking the limit $\eta\rightarrow\inf$. Therefore, in the large $\eta$ limit, we find
\be
\mF_n(\m,x)=\frac{s_{n,\m}}{\prod_{k=0}^{n-1}F_a(1-x)F_a(x)},
\ee
which recovering the correct quantum result with the proper $\eta$ dependent normalisation.
\section{Charged R\'enyi negativity}\label{section5}
As explained in section \ref{section2}, for the charged R\'enyi negativity defined in eq.~(\ref{Rnmu}), we have the following relation
\be
\begin{split}
&R_n(\m)=\langle\mathcal{T}_a(u_1)\tilde{\mathcal{T}}_a(v_1)\tilde{\mathcal{T}}_a(u_2)\mathcal{T}_a(v_2)\rangle\\
&=c_{n,\m}^2\left(\frac{(u_2-u_1)(v_2-v_1)}{(v_1-u_1)(v_2-u_2)(v_2-u_1)(u_2-v_1)}\right)^{2\D_{n,\m}}\mathcal{G}_{n}(\m;y),
\end{split}
\ee
where we have introduced the ordered four point ratio
\be
y=\frac{(v_1-u_1)(v_2-u_2)}{(u_2-u_1)(v_2-v_1)}=\frac{x}{x-1}
\ee
with $0<y<1$. The two scaling function $\mF_{n}(\m;x)$ and $\mG_n(\m;y)$ are related as
\be
\mG_n(\m;x)=(1-y)^{4\D_{n,\m}}\mF_n\lt(\m;\frac{y}{y-1}\rt).
\ee
Since the scaling function $\mF_{n}(\mu,x)$ given in eq.~(\ref{Fnmux}) is only valid in the region $0<x<1$, while in our case $-\inf<\frac{y}{y-1}<0$. So we must find the expression of the scaling function $\mF_n(\m;x,\bar x)$ for generic four-point ratio $x,\bar x\in\mathbb{C}$. As explained in appendix \ref{AppendixA}, for generic $x\in\mathbb{C}$, $\mF_n(\m;x,\bar x)$ reads
\be
\mF_n(\m;x,\bar x)=\prod_{k=0}^{n-1}\frac{\mathrm{const}}{F_a(x)\bar F_a(1-\bar x)+\bar F_a(\bar x)F_a(1-x)}\sum_{\mathbf{m}\in\mathbb{Z}^{2n}}\prod_{k=0}^{n-1}e^{-S_a^{\mathrm{cl}}}.
\ee
In this case, the classical action is given by eq.~(\ref{ClassicalAction})
\be\label{action1}
S_a^{\mathrm{cl}}=\frac{2g\pi\sin(\pi a)}{n}\lt[\frac{|\tau_a|^2}{\beta_a}|\xi_1|^2+\frac{\a_a}{\beta_a}(\xi_1\bar\xi_2\bar\g+\bar\xi_1\xi_2\g)+\frac{|\xi_2|^2}{\beta_a}\rt].
\ee
For generic $x\in\mathbb{C}$, the real part $\a_a(x)$ of the modulus $\tau_a(x)$ of a ``fake torus" is nonzero
\be
\tau_a(x)\equiv\mathrm{i}\frac{F_a(1-x)}{F_a(x)}=\a_a(x)+\mathrm{i}\b_a(x).
\ee
Substitute the explicit expression of $\xi_i$ into eq.~(\ref{action1}) and summing over $k$ and $\mathbf{m}$, the classical part of the partition function can be written as some Siegel theta function. This calculation is almost the same with \cite{calabrese2013entanglement}, and the final result can be obtained by tiny adjustment
\be\label{Fnmu}
\mF_n(\m;x,\bar x)=s_{n,\m}\frac{\Theta(\mathbf{0}|\eta G)^2}{\prod_{k=0}^{n-1}\Re[F_a(x)\bar F_a(1-\bar x)]}.
\ee
The $2n\times 2n$ symmetric matrix $G$ introduced in the $\mathrm{r.h.s}$ of eq.~(\ref{Fnmu}) is purely imaginary and can be written in terms of $n\times n$ block matrices as
\be
G=2\mathrm{i}\begin{pmatrix}
A&W\\
W^{\mathrm{t}}&B
\end{pmatrix},
\ee
with
\be
\begin{split}
&A_{rs}=\frac{1}{n}\sum_{k=0}^{n-1}\frac{|\tau_a|^2}{\b_a}\sin(\pi a)\cos\Big[\frac{2\pi k}{n}(r-s)\Big],\\
&B_{rs}=\frac{1}{n}\sum_{k=0}^{n-1}\frac{1}{\b_a}\sin(\pi a)\cos\Big[\frac{2\pi k}{n}(r-s)\Big],\\
&W_{rs}=-\frac{1}{n}\sum_{k=0}^{n-1}\frac{\a_a}{\b_a}\sin(\pi a)\sin\Big[\frac{2\pi k}{n}(r-s+\frac12)\Big].
\end{split}
\ee
For $x\in(0,1)$, we have $\a_a(x)=0$ for every $k$ and therefore the off diagonal blocks $W$ of $G$ are zero, then the Siegel theta function $\Theta(\mathbf{0}|\eta G)$ factorizes into a product of two Siegel theta functions $\Theta(\mathbf{0}|\eta A)\Theta(\mathbf{0}|\eta B)=\Theta(\mathbf{0}|\eta\O)\Theta(\mathbf{0}|\eta\td{\O})$ and eq.~(\ref{Fnmu}) gives back eq.~(\ref{Fnmux}).
\par Then we can write $\mG(\m;y)$ for $0<y<1$ as
\be
\mG(\m;y)=s_{n,\m}(1-y)^{4\D_{n,\m}}\frac{\Theta(\mathbf{0}|\eta G(\frac{y}{y-1}))^2}{\prod_{k=0}^{n-1}\Re[F_a(\frac{y}{y-1})\bar F_a(\frac{1}{1-y})]}.
\ee
In the decompactification limit $\eta\rightarrow\inf$, we have
\be
\mG(\m;y)=\frac{s_{n,\m}(1-y)^{4\D_{n,\m}}}{\prod_{k=0}^{n-1}\Re[F_a(\frac{y}{y-1})\bar F_a(\frac{1}{1-y})]}.
\ee
\section{Numerical results }\label{section6}
In this section, we consider the charged moments and the charged R\'enyi negativity for the complex harmonic chain with periodic boundary conditions. We will use this lattice model to check the CFT formula obtained in section \ref{section4} and section \ref{section5} in the decompactification regime.
\par The Hamiltonian of the real harmonic chain made by $L$ sites reads
\be
H_{HC}=\sum_{j=0}^{L-1}\lt(\frac{1}{2M}p_j^2+\frac{M\o^2}{2}q_j^2+\frac{K}{2}(q_{j+1}-q_j)^2\rt),
\ee
where periodic boundary conditions $q_L\equiv q_0,p_L\equiv p_0$ are imposed and variables $p_j$ and $q_j$ satisfy standard bosonic commutation relations $[q_i,q_j]=[p_i,p_j]=0$ and $[q_i,p_j]=\mathrm{i}\d_{ij}$. We can work with $M=K=1$ without loss of generality. The lattice version of the complex non-compact bosonic field theory is the sum of two of the above harmonic chain. In terms of the variables $q^{(1)},p^{(1)}$ and $q^{(2)},p^{(2)}$, the Hamiltonian is
\be\label{HCHC}
H_{CHC}(p^{(1)}+\mathrm{i}p^{(2)},q^{(1)}+\mathrm{i}q^{(2)})=H_{HC}(p^{(1)},q^{(1)})+H_{HC}(p^{(2)},q^{(2)}).
\ee
\par Since the bosonic field is not compactified and massless, we must compare the continuum limit of eq.~(\ref{HCHC}) with the regime $\eta\rightarrow\inf$ of the CFT results computed in section \ref{section4} and section \ref{section5}.
The Hamiltonian eq.~(\ref{HCHC}) can be diagonalized by introducing the creation and annihilation operators $a_k,a_k^{\dg}$ and $b_k,b_k^{\dg}$, satisfying $[a_k,a_{k'}^{\dg}]=\d_{kk'}$ and $[b_k,b_{k'}^{\dg}]=\d_{kk'}$. In terms of these operators, the Hamiltonian eq.~(\ref{HCHC}) is diagonal
\be
H_{CHC}=\sum_{k=0}^{L-1}\e_k(a_k^{\dg}a_k+b_k^{\dg}b_k),\quad \e_k=\sqrt{\mathbf{\o}^2+\frac{4K}{M}\sin^2\lt(\frac{\pi k}{L}\rt)}.
\ee
While the $U(1)$ charge is
\be
Q=\sum_{k=0}^{L-1}\e_k(a_k^{\dg}a_k-b_k^{\dg}b_k).
\ee
The conserved charge is local and can also be written in the position space and for a given subsystem $A$ reads
\be
Q_A=\sum_{j\in A}\e_k(a_j^{\dg}a_j-b_j^{\dg}b_j).
\ee
\subsection{Charged moments}
\begin{figure}
        \centering
        \subfloat
        {\includegraphics[width=7cm]{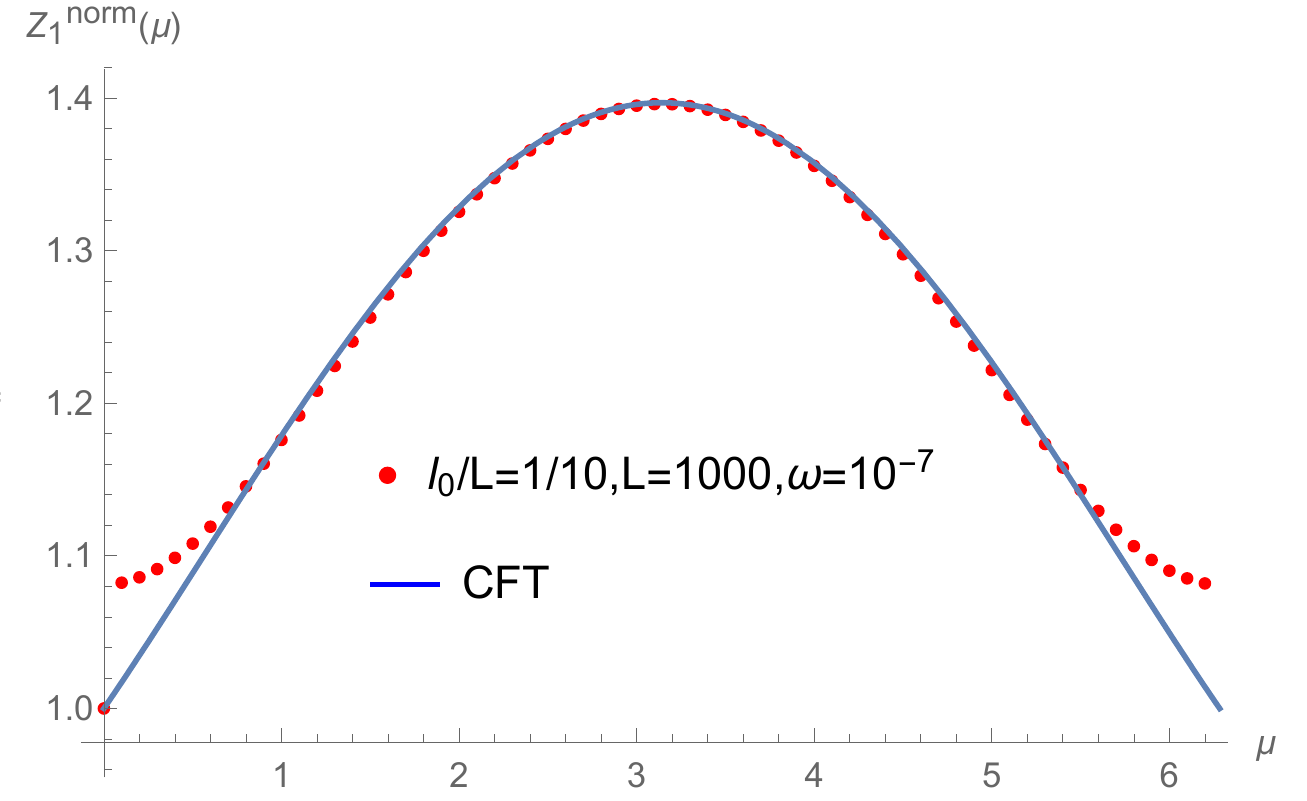}} \quad\quad
        {\includegraphics[width=7cm]{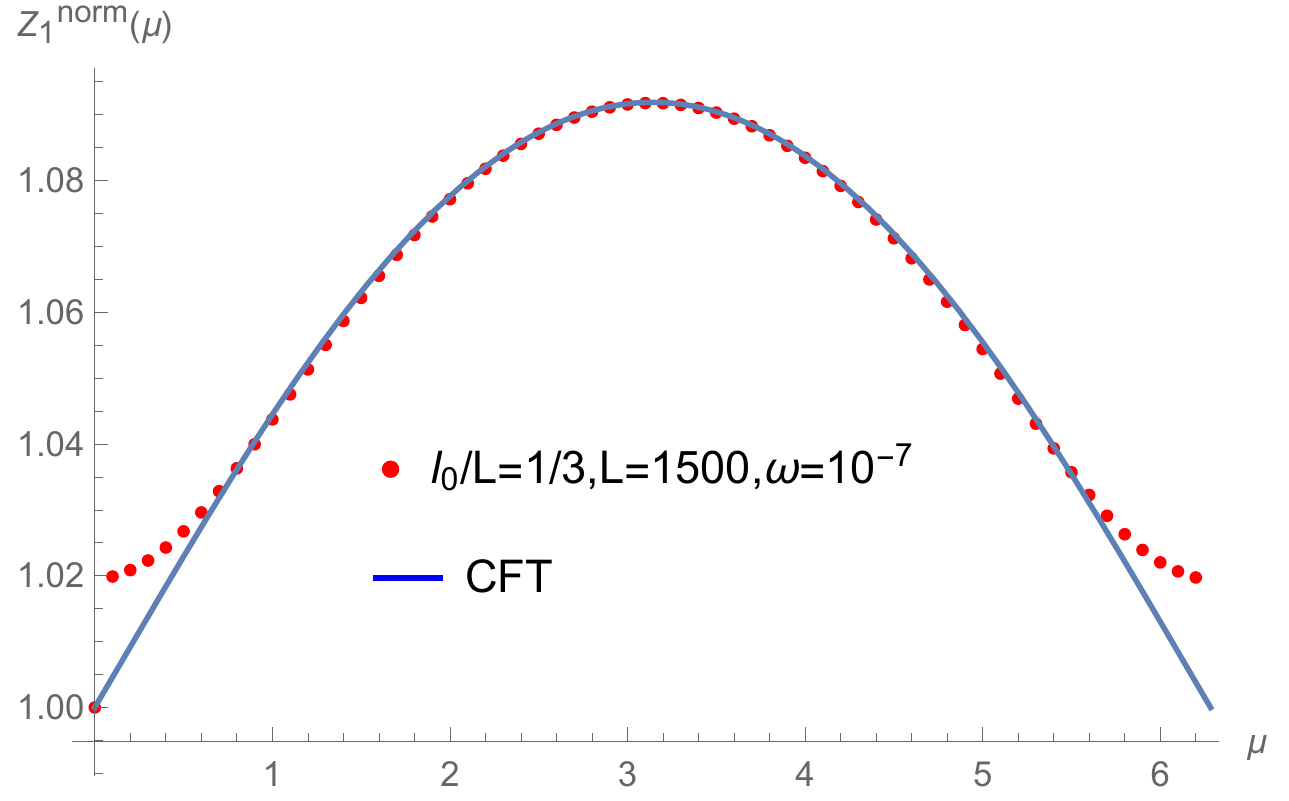}}
        \caption{Numerical data of $Z_1^{\text{norm}}$ as a function of $\m$ for different $z=l_0/L$ in the complex harmonic chain. The full lines are the CFT predictions. Here we consider $l_0=100,d=400,L=1000,\o=10^{-7}$ and $l_0=500,d=250,L=1500,\o=10^{-7}$. The agreement is very well for $\m$ near $\pi$, but it worsens as $\m$ gets closer to $0$ and $2\pi$.}
        \label{fig2}
\end{figure}

The charged moments factorise as
\be
Z_{n}(\m)=\Tr\rho_A^ne^{\mathrm{i}\m Q_A}=\Tr[\rho_A^ne^{\mathrm{i}\m N_A^a}]\times\Tr[\rho_A^ne^{-\mathrm{i}\m N_A^b}].
\ee
The correlation functions for real harmonic chain are
\be
\begin{split}
\mathbf{Q}_{rs}\equiv\bra{0}q_rq_s\ket{0}=\frac{1}{2L}\sum_{k=0}^{L-1}\frac{1}{M\o_k}\cos\lt(\frac{2\pi k(r-s)}{L}\rt),\\
\mathbf{P}_{rs}\equiv\bra{0}p_rp_s\ket{0}=\frac{1}{2L}\sum_{k=0}^{L-1}M\o_k\cos\lt(\frac{2\pi k(r-s)}{L}\rt).
\end{split}
\ee
when $r,s=0,1,\cdots,L-1$ run over the whole chain, $\mathbf{Q}$ and $\mathbf{P}$ are $L\times L$ matrix, satisfy $\mathbf{Q}\mathbf{P}=\frac14\mathbf{I}$. The limit $\o\rightarrow 0$ is ill defined since zero mode in $\mathbf{Q}_{rs}$ diverges. Therefore to test our analytic result, we must keep $\o>0$ and let $\o L\ll 1$ in order to stay in the conformal regime.
\begin{figure}
        \centering
        \subfloat
        {\includegraphics[width=5cm]{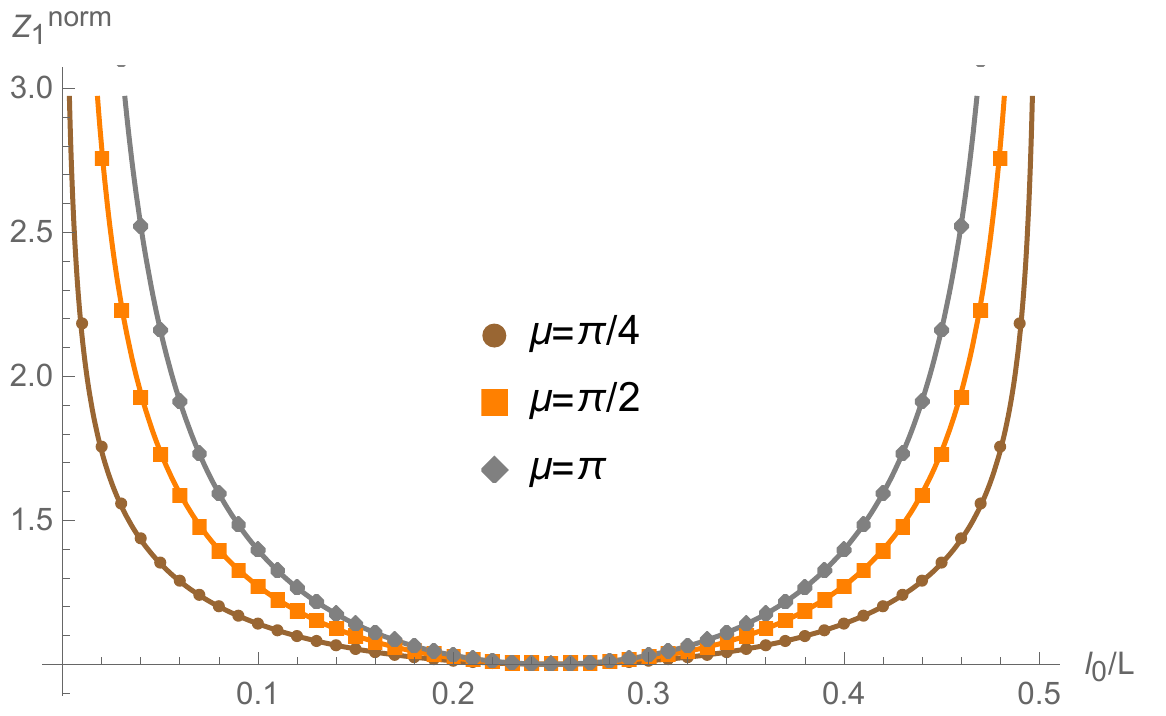}} \quad
        {\includegraphics[width=5cm]{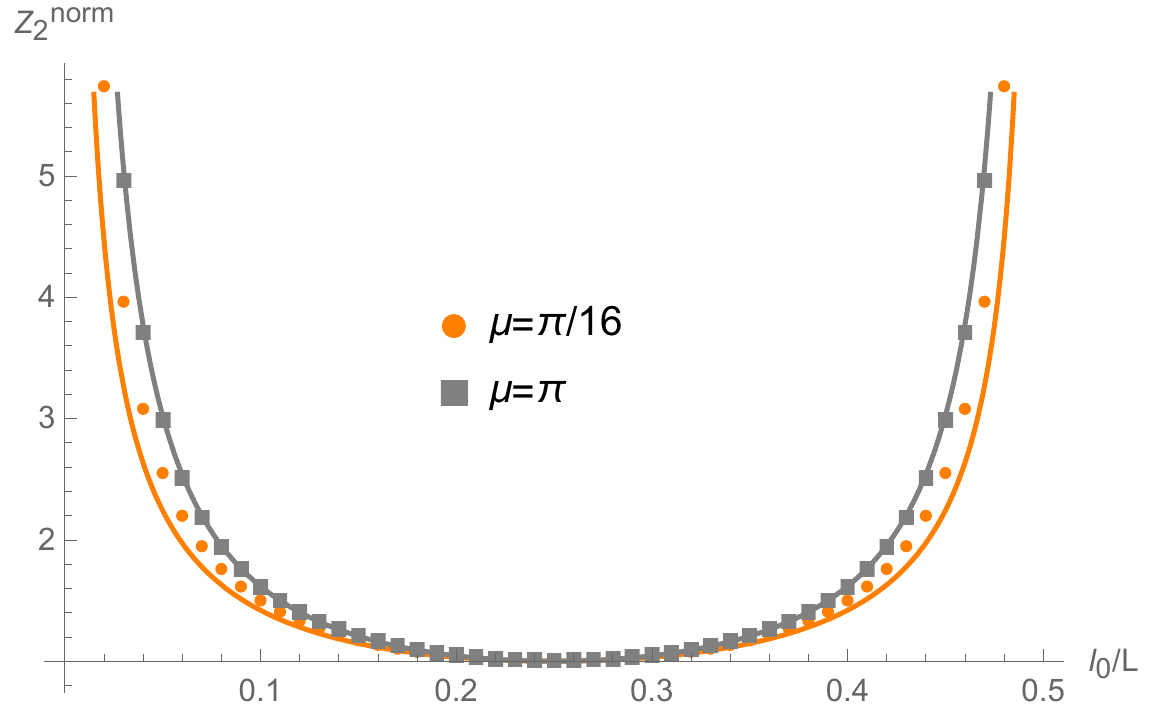}}\quad
        {\includegraphics[width=5cm]{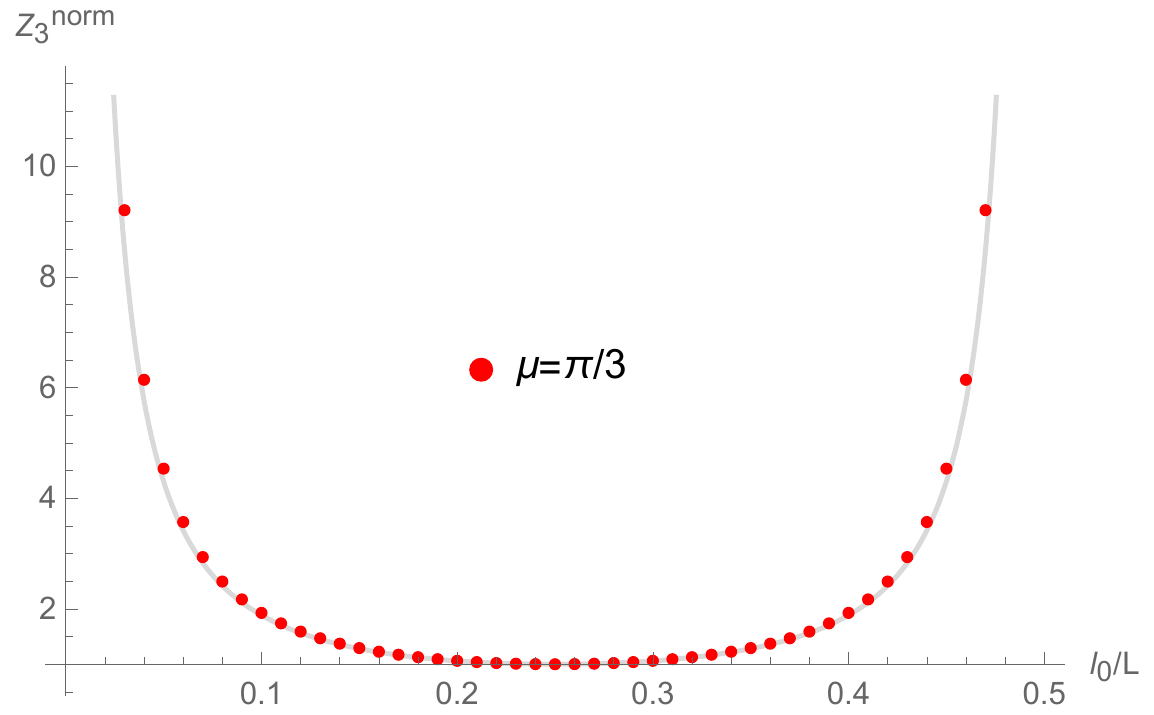}}
        \caption{Numerical data of $Z_n^{\text{norm}}$ as a function of $z=l_0/L$ for different $n$ and $\m$ in the complex harmonic chain. The full lines are the CFT predictions (cf. eq.~(\ref{ZnCFT})). Here we consider $\o=10^{-8},L=2000$ and $\m=\frac{\pi}{4},\frac{\pi}{2},\pi$ for $n=1$, $\m=\frac{\pi}{16},\pi$ for $n=2$ and $\m=\frac{\pi}{3}$ for $n=3$. As shown in the figure, the agreement is very well.}
        \label{fig3}
\end{figure}

\par The charged R\'enyi entropies of any subsystem $A$ consist of $l$ lattice sites can be computed from these correlators. Firstly, we have to consider the correlation matrices $\mathbf{Q}_A$ and $\mathbf{P}_A$ obtained by restricting the the indices of the correlation matrices $\mathbf{Q}$  and $\mathbf{P}$ to the sites belonging to $A$. Then find the eigenvalues of the $l\times l$ matrix $\mathbf{Q}_A\mathbf{P}_A$ which are denoted by $\{\s_1^2,\cdots,\s_l^2\}$. We also introduce the basis $\ket{\mathbf{n}}\equiv\otimes_{k=1}^l\ket{n_k}$, defined by products of Fock states of the number operator in the subsystem $A$, the reduced density matrix of $A$ can be written as
\be
\rho_A=\sum_{\mathbf{n}}\prod_{k=1}^{l}\frac{1}{\s_k+1/2}\lt(\frac{\s_k-1/2}{\s_k+1/2}\rt)^{n_k}\ket{\mathbf{n}}\bra{\mathbf{n}}.
\ee
Finally, the charged moments for real harmonic chain are given by
\be
\begin{split}
\Tr\rho_A^ne^{\mathrm{i}\m N_A}&=\sum_{\mathbf{n}}\prod_{k=1}^{l}\lt[\frac{1}{\s_k+1/2}\lt(\frac{\s_k-1/2}{\s_k+1/2}\rt)^{n_k}\rt]^ne^{\mathrm{i}n_k\m}\\
&=\prod_{k=1}^l\lt[\lt(\s_k+\frac12\rt)^n-e^{\mathrm{i}\m}\lt(\s_k-\frac12\rt)^n\rt]^{-1}.
\end{split}
\ee
Therefore the charged moments of RDM of an arbitrary subsystem $A$ for a complex harmonic chain are
\be
\begin{split}
&Z_n(\m)=|\Tr\rho_A^ne^{\mathrm{i}\m N_A}|^2\\
&=\prod_{k=1}^l\lt[\lt(\s_k+\frac12\rt)^n-e^{\mathrm{i}\m}\lt(\s_k-\frac12\rt)^n\rt]^{-1}\lt[\lt(\s_k+\frac12\rt)^n-e^{-\mathrm{i}\m}\lt(\s_k-\frac12\rt)^n\rt]^{-1}.
\end{split}
\ee
This method also hold when $A$ is the union of two disjoint intervals $A_1,A_2$, which is the situation we consider in the CFT approach.
\par Let us denote the number of sites in $A_1$ and $A_2$ by $l_1$ and $l_2$ respectively, and denote the number of sites in the separation between $A_1$ and $A_2$ by $d_1$ and $d_2$ in the periodic chain. Then $l=l_1+l_2$ and $L=l_1+l_2+d_1+d_2$ must be imposed.
For simplicity, we will consider the configuration where all the intervals have the same length and all the separations have the same size, namely
\be
l_1=l_2\equiv l_0,\quad d_1=d_2\equiv d.
\ee
In this situation, we have only one free parameter $l_0/L$ to character the configuration since $d/L=1/2-l_0/L$.
\par To compare our numerical data with the CFT results obtained in the previous sections, we must map the CFT formulas on the complex plane to the cylinder of circumference $L$. This can be easily done by replace each length $y$ (e.g. $l_0,d,l_0+d$,etc.) with the corresponding chord length $\frac{L}{\pi}\sin\frac{\pi y}{L}$. For the cross ratio
\be
x=\frac{\sin^2(\pi l_0/L))}{\sin^2(\pi(l_0+d)/L)}=\sin^2\lt(\frac{\pi l_0}{L}\rt).
\ee
For the CFT on a cylinder of circumference $L$, we have
\be\label{ZnCFT}
\begin{split}
Z_n(\m,z)=&c_{n,\m}^2s_{n,\m}\left(\frac{\pi^2}{L^2\sin^2(\pi z)\sin[\pi(z+1/2)]\sin[\pi(1/2-z)]}\right)^{4\D_{n,\m}}\\
&\times\prod_{k=0}^{n-1}\left[F_{\frac{k}{n}+\frac{\m}{2\pi n}}(\sin^2(\pi z))F_{\frac{k}{n}+\frac{\m}{2\pi n}}(1-\sin^2(\pi z))\right]^{-1},
\end{split}
\ee
where $z\equiv \frac{|u_1-v_1|}{L}=\frac{|u_2-v_2|}{L}$.
In order to eliminate the unknown parameter $c_{n,\m}$, it is more convenient to normalize the results through a fixed configuration. We choose the following one
\be
\text{fixed configuration}:\quad l_1=l_2=d_1=d_2=\Big[\frac{L}{4}\Big]
\ee
and denote the corresponding charged moments by $Z_{n}^{\text{fixed}}(\m)$.
\par In the remaining part, we assume that the length of chain $L$ multiples of 4 for simplicity. Then we have $Z_{n}^{\text{fixed}}(\m)=Z_n(\m,1/4)$.
The numerical data and the CFT results shown in the figures are compared in terms of their normalised version
\be
Z_n^{\text{norm}}(\m,z)=\frac{Z_n(\m,z)}{Z_{n}^{\text{fixed}}(\m)}=[\csc(2\pi z)]^{4\D_{n,\m}}\prod_{k=0}^{n-1}\frac{[F_{\frac{k}{n}+\frac{\m}{2\pi n}}(1/2)]^2}{F_{\frac{k}{n}+\frac{\m}{2\pi n}}(\sin^2(\pi z))F_{\frac{k}{n}+\frac{\m}{2\pi n}}(1-\sin^2(\pi z))}.
\ee
In the limit $L\rightarrow\inf$ with $\o L$ kept fixed, our numerical results should converge to the CFT computations for $Z^{\mathrm{norm}}_n(\m,x)$  in the decompactification regime.
\par The numerical results for the function $Z_1^{\mathrm{norm}}(\m)$  are reported in Fig.~\ref{fig2} for different $L$ and different sub subsystem sizes $l_0$. As shown in the figure, the agreement between numerical data and CFT prediction is excellent for $\m$ near $\pi$, while it gets worse for $\m$ close to $0$ and $2\pi$. We found that the convergence to the CFT results is not uniform and it's much more slower for $\m$ near $0$ and $2\pi$. This can be explained by the fact that the limit $\m\rightarrow 0$ or $\m\rightarrow 2\pi$ does not commute with the critical limit $\o\rightarrow 0$ \cite{Murciano:2019wdl, Murciano:2020vgh}.
\par In Fig.~\ref{fig3}, we report the numerical data for the quantities $Z_n^{\mathrm{norm}}(\m,z)$ for various $n$ and $\m$. As the figure shows, the numerical results and the CFT predictions match very well.
\begin{figure}
        \centering
        \subfloat
        {\includegraphics[width=7cm]{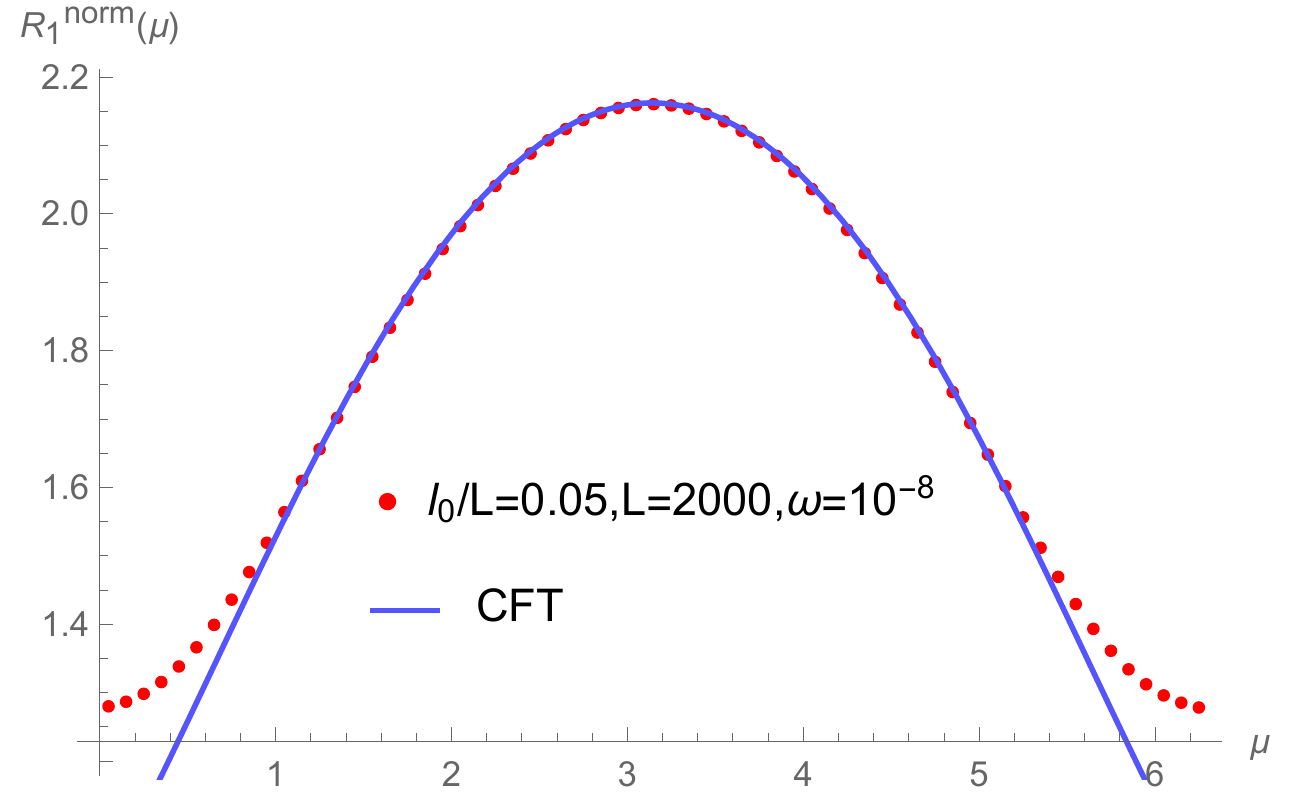}} \quad\quad
        {\includegraphics[width=7cm]{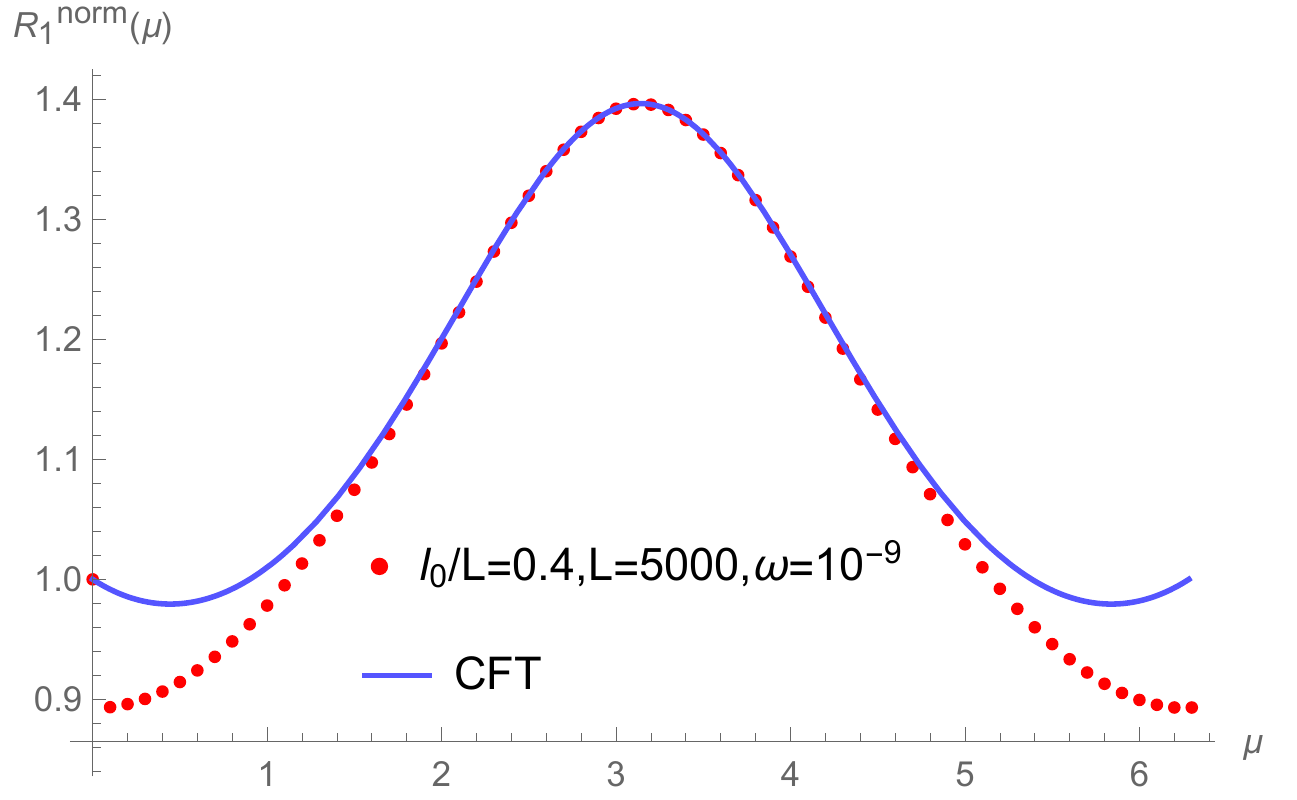}}
        \caption{Numerical data of $R_1^{\text{norm}}$ as a function of $\m$ for different $z=l_0/L$ in the complex harmonic chain. The full lines are the CFT predictions (cf. eq.~(\ref{RnCFT})). Here we consider $l_0=100,L=2000,\o=10^{-8}$ and $l_0=2000,L=5000,\o=10^{-9}$. Again, the agreement is very well for $\m$ near $\pi$, but it worsens as $\m$ gets closer to $0$ and $2\pi$.}
        \label{fig4}
\end{figure}
\subsection{Charged R\'enyi negativity}
In the Fock basis $\{\ket{\mathbf{n}}\}$, $Q_2^{T_2}=Q_2$ and the operator $\mQ_A=Q_1-Q^{T_2}_2=Q_1-Q_2$ becomes exactly the charge imbalance operator. For the complex harmonic chain, the charged R\'enyi negativity factorises as
\be
R_{n}(\m)=\Tr[(\rho^{T_2}_A)^ne^{\mathrm{i}\m (Q_1-Q_2)}]=\Tr[(\rho^{T_2}_A)^ne^{\mathrm{i}\m \mQ_A^a}]\times\Tr[(\rho^{T_2}_A)^ne^{-\mathrm{i}\m \mQ_A^b}].
\ee
In bosonic system, the net effect of partial transposition with respect to $A_2$ is changing the sign of the momenta corresponding to $A_2$. Thus the momenta correlators in the partial transposed density matrix can be obtained from $\mathbf{P}_A$ by simply change the sign of the momenta that in $A_2$, i.e.
\be
\mathbf{P}_A^{T_2}=\mathbf{R}_{A_2}\mathbf{P}_A\mathbf{R}_{A_2},
\ee
where $\mathbf{R}_{A_2}$ is the $l_0\times l_0$ diagonal matrix with elements $(\mathbf{R}_{A_2})_{rs}=(-1)^{\d_{r\in A_2}}\d_{rs}$. We denote the eigenvalues of $\mathbf{Q}_A\mathbf{P}_A^{T_2}$ by $\{\tau_1^2,\tau_2^2,\cdots,\tau_l^2\}$.
Then we have
\be
\Tr[(\rho^{T_2}_A)^ne^{\mathrm{i}\m \mQ_A^a}]=\prod_{j= 1}^{l_0}\lt[\lt(\t_j+\frac12\rt)^n-e^{\mathrm{i}\m}\lt(\t_j-\frac12\rt)^n\rt]^{-1}\prod_{k= l_0+1}^{l}\lt[\lt(\t_k+\frac12\rt)^n-e^{-\mathrm{i}\m}\lt(\t_k-\frac12\rt)^n\rt]^{-1}.
\ee
As before, in the large $L$ limit while with $\o L$ kept fixed, our numerical results should converge to the CFT computations for $R^{\mathrm{norm}}_n(\m,x)$  in the decompactification regime.
\par For the non-compact free boson CFT on a cylinder of circumference $L$, we have
\be\label{RnCFT}
R_n^{\text{norm}}(\m,z)=\frac{R_n(\m,z)}{R_{n}^{\text{fixed}}(\m)}=[\cot(\pi z)]^{4\D_{n,\m}}\prod_{k=0}^{n-1}\frac{\Re[F_{\frac{k}{n}+\frac{\m}{2\pi n}}(-1)\bar{F}_{\frac{k}{n}+\frac{\m}{2\pi n}}(-2)]}{\Re[ F_{\frac{k}{n}+\frac{\m}{2\pi n}}(\frac{\sin^2(\pi z)}{\sin^2(\pi z)-1})\bar{F}_{\frac{k}{n}+\frac{\m}{2\pi n}}(\frac{1}{1-\sin^2(\pi z)})]}
\ee
and
\be\label{GnCFT}
\mG_n^{\text{norm}}(\m,z)=[\cos^2(\pi z)]^{-4\D_{n,\m}}R_n^{\text{norm}}(\m,z).
\ee
While for the harmonic chain, we have
\be\label{Gnlat}
\mG_{n,\text{lat}}^{\text{norm}}(\m,l_0/L)=\left[4\sin\left[\pi\left(\frac12-\frac{l_0}{L}\right)\right]\sin\left[\pi\left(\frac12+\frac{l_0}{L}\right)\right]
\sin^2\left(\frac{\pi l_0}{L}\right)\right]^{4\D_{n,\m}}\Tr[(\rho^{T_2}_A)^ne^{\mathrm{i}\m\mQ_A}].
\ee
\par The numerical results for the function $R_1^{\mathrm{norm}}(\m)$  are reported in Fig.~\ref{fig4} for different $L$ and different sub subsystem sizes $l_0$. As shown in the figure, the agreement between numerical data and CFT prediction is pretty well for $\m$ near $\pi$, while it gets worse for $\m$ close to $0$ and $2\pi$ as before.
\par
\begin{figure}
        \centering
        \subfloat
        {\includegraphics[width=7cm]{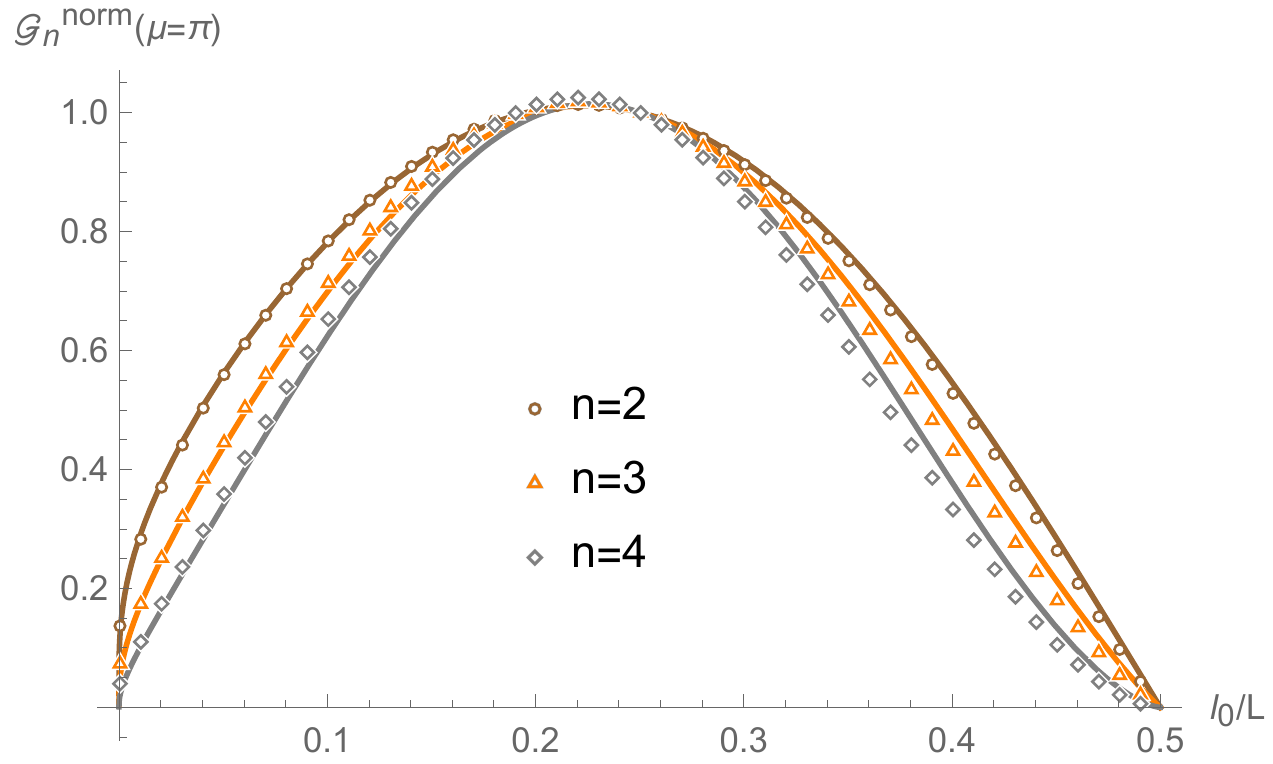}} \quad\quad
        {\includegraphics[width=7cm]{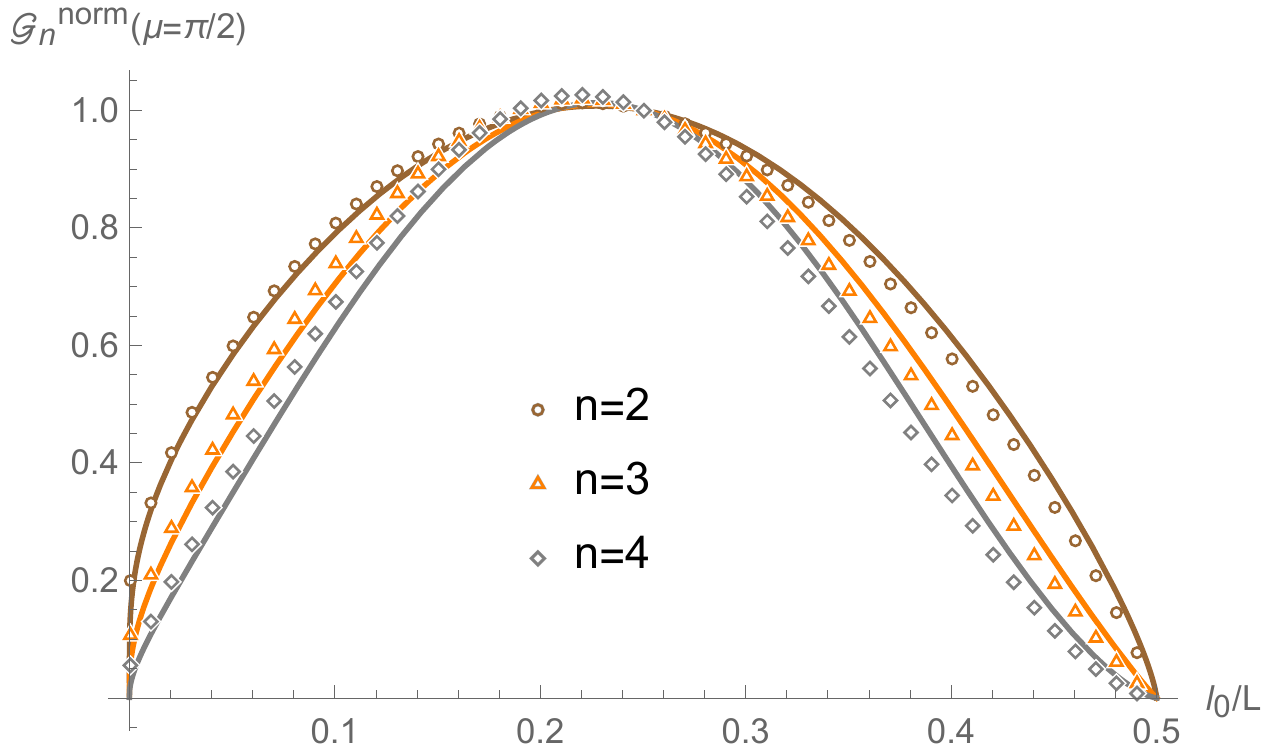}}
        \caption{Numerical data of $\mG_n^{\text{norm}}$ as a function of $z=l_0/L$ for different $n$ and $\m$ in the complex harmonic chain(cf. eq.~(\ref{Gnlat})). The full lines are the CFT predictions (cf. eq.~(\ref{GnCFT})). Here we consider $L=2000,\o=10^{-9}$ for $n=2,3,4$ with $\m=\pi$ and $\m=\pi/2$ respectively. The agreement is very well for small $n$, but it worsens as $n$ goes larger.}
        \label{fig5}
\end{figure}

 In Fig.~\ref{fig5}, we report the numerical data for the quantities $\mG_n^{\mathrm{norm}}(\m,x)$ for various $n$ and $\m$. In this case, as shown in the figure, the numerical results and the CFT predictions matched excellently.
\section{Conclusion and discussion}\label{section7}
In this paper, we consider the charged moments of reduced density matrices of two disjoint regions in 1+1 dimensional compact free boson CFT. We discuss two important regimes of the scaling function, one is the small $x$ regime, which will be useful to study the conformal block of the fluxed twist fields. The other region is the decompactification regime, in this regime, the bosonic fields become non-compact, and will be convenient for numerical checks. We also compute the charged R\'enyi negativity.
We test our analytic results against exact numerical computations in the complex harmonic chain, finding perfect agreement.
\par We must mention we just obtained the charged moments and the charged R\'enyi negativity, it would be very interesting to apply Fourier transformation and analytic continue our results to obtain the corresponding symmetry resolved entanglement entropy and charge imbalance resolved entanglement negativity.
\par It would be also interesting the further check our analytic formula by considering the conformal block expansion of the four-point correlation function of fluxed twist fields and applying the Zamolodchikov recursion formula \cite{zamolodchikov1984conformal, Ruggiero:2018hyl} for each conformal block. In this method, each term in the conformal block expansion can be analytically continued and this approach can provide a good approximation for the symmetry resolved entanglement entropy and charge imbalance resolved entanglement negativity.
\par In this paper, we focus on the charged moments and charged R\'enyi negativity of complex boson with $U(1)$ symmetry. In 1+1 dimensional, for fermions with $U(1)$ symmetry $\psi_k\rightarrow e^{\mathrm{i}\m}\psi_k$, the corresponding fluxed twist field admits a bosonisation formula. We can write the complex fermionic field as $\psi_k\sim e^{\mathrm{i}\varphi_k}$, then the fermion twist field is given by the vertex operator $\mT_{n,k,\m}=e^{\mathrm{i}(\frac{k}{n}+\frac{\m}{2\pi n})\varphi_k}$, which have a different scaling dimension $\D_{n,k,\m}=\left(\frac{k}{n}+\frac{\m}{2\pi n}\right)^2$. This can be understood, since via bosonisation of $U(1)$
complex fermions, the corresponding bosons transform by translation, and should instead satisfy boundary condition $\varphi_k(e^{2\pi\mathrm{i}} z)=\varphi_k(z)+\m$.
\section*{Acknowledgments}
This work was supported  by the National Natural Science Foundation of China, Grant No.\ 12005081.
\begin{appendix}
\section{Correlation function of fluxed twist fields}\label{AppendixA}
In this section, we give some detail of the calculation of four point correlation functions of the $U(1)$ twist fields $\mT_{n,k,\m}$. We will call it generally $\mT_a$, where $a=\frac{k}{n}+\frac{\m}{2\pi n}$ with $0<a<1$. This problem is already encountered in the theory of orbifold CFT \cite{dixon1987conformal}. Here we can just borrow the results with slightly change. However, for the reader's convenience we briefly review this method here.
\par Let us consider a complex scalar field $X(z,\bar z)$ defined on a Riemann sphere. The action is given by
\be
\mS[X,\bar X]=\frac{1}{4\pi}\int d^2z(\p X\bar\p\bar X+\bar\p X\p\bar X)
\ee
If the field $\mT_{n,\m,k}$ inserted at the origin, then we have
\be
X(e^{2\pi\mathrm{i}}z,e^{-2\pi\mathrm{i}}\bar z)=e^{2\pi\mathrm{i}a} X(z,\bar z)+\l,
\ee
where $\l\in \L$. Let us now split the field $X$ into a classical piece $X_{\mathrm{cl}}$ and a quantum fluctuation $X_{\mathrm{qu}}$ as described in section \ref{section4}. The classical field should behave the same as the full field
\be
X_{\mathrm{cl}}(e^{2\pi\mathrm{i}}z,e^{-2\pi\mathrm{i}}\bar z)=e^{2\pi\mathrm{i}a} X_{\mathrm{cl}}(z,\bar z)+\l,
\ee
which implies
\be
X_{\mathrm{qu}}(e^{2\pi\mathrm{i}}z,e^{-2\pi\mathrm{i}}\bar z)=e^{2\pi\mathrm{i}a} X_{\mathrm{qu}}(z,\bar z),
\ee
In order to determine the quantum Green functions in the presence of the twist fields and classical solutions, one needs some global information that can be obtained from the monodromy conditions for transporting $X_{\mathrm{qu}}$ and $X_{\mathrm{cl}}$ around collections of twist fields that have net twist zero. Such paths $\mC_i$ are called closed loops. For any closed loop $\mC_i$, we must have
\be\label{Monodromyqu}
\D_{\mC_i}X_{\mathrm{qu}}=\oint_{\mC_i}dz~\partial X_{\mathrm{qu}}+\oint_{\mC_i}d\bar z~\bar\partial X_{\mathrm{qu}}=0
\ee
and
\be
\D_{\mC_i}X_{\mathrm{cl}}=\oint_{\mC_i}dz~\partial X_{\mathrm{cl}}+\oint_{\mC_i}d\bar z~\bar\partial X_{\mathrm{cl}}=\l,
\ee
where now for our case $\l\in(1-e^{2\pi\mathrm{i}a})\L$.
\par Now let us consider the following four-point function of the twist field $\mT_a$
\be
Z=\langle\mT_a(z_1,\bar z_1)\td{\mT}_a(z_2,\bar z_2)\mT_a(z_3,\bar z_3)\td{\mT}_a(z_4,\bar z_4)\rangle=\int[dX][d\bar X]e^{-\mS[X,\bar X]}=Z_{\mathrm{qu}}\sum_{X_{\mathrm{cl}}}e^{-\mS[X_{\mathrm{cl}},\bar X_{\mathrm{cl}}]}.
\ee
We can first compute the Green function in the presence of four twist fields
\be
g(z,w;z_i)=\frac{\langle-\frac12\p_z X\p_w\bar X\mT_a(z_1)\td{\mT}_a(z_2)\mT_a(z_3)\td{\mT}_a(z_4)\rangle}{\langle\mT_a(z_1)\td{\mT}_a(z_2)\mT_a(z_3)\td{\mT}_a(z_4)\rangle}.
\ee
This Green function should obey the asymptotic conditions: $g(z,w;z_i)\sim (z-w)^{-2}$ as $z\rightarrow w$ and for $z\rightarrow z_i$ we have
$g(z,w;z_i)\sim (z-z_j)^{-a}$ when $j=1,3$ and $g(z,w;z_i)\sim (z-z_j)^{-(1-a)}$ when $j=2,4$.
The holomorphic fields for the cut $z$-plane in this case are:
\be
\begin{split}
\p X(z)\equiv\o_a(z)=[(z-z_1)(z-z_3)]^{-a}[(z-z_2)(z-z_4)]^{-(1-a)},\\
\p \bar X(z)\equiv\o_{1-a}(z)=[(z-z_1)(z-z_3)]^{-(1-a)}[(z-z_2)(z-z_4)]^{-a}.
\end{split}
\ee
Then the unique form of the Green function with the desired properties reads
\be
\begin{split}
g(z,w;z_i)=\o_a(z)\o_{1-a}&(w)\Big[a\frac{(z-z_1)(z-z_3)(w-z_2)(w-z_4)}{(z-w)^2}\\
&+(1-a)\frac{(z-z_2)(z-z_4)(w-z_1)(w-z_3)}{(z-w)^2}+A(z_i,\bar z_i)\Big].
\end{split}
\ee
We can find the expectation value of the stress tensor $T=-\frac12:\p X\p\bar X:$ in the presence of twist fields by taking $w\rightarrow z$ of $g(z,w;z_i)$ and subtracting the diverging pieces. We have
\be
\begin{split}
&\frac{\langle T(z)\mT_a(z_1)\td{\mT}_a(z_2)\mT_a(z_3)\td{\mT}_a(z_4)\rangle}{\langle\mT_a(z_1)\td{\mT}_a(z_2)\mT_a(z_3)\td{\mT}_a(z_4)\rangle}=\lim_{w\rightarrow z}(g(z,w;z_i)-(z-w)^{-2})\\
&=\frac12a(1-a)\lt(\frac{1}{z-z_1}+\frac{1}{z-z_3}-\frac{1}{z-z_2}-\frac{1}{z-z_4}\rt)^2+\frac{A}{(z-z_1)(z-z_2)(z-z_3)(z-z_4)}.
\end{split}
\ee
From the operator product expansion
\be
T(z)\td{\mT}_a(z_2)\sim \frac{h_a\td{\mT}_a(z_2)}{(z-z_2)^2}+\frac{\p_{z_2}\td{\mT}_a(z_2)}{z-z_2},
\ee
it's easy to know that the conformal dimensions of the twist fields $\td{\mT}_a$ (and also of $\mT_a$) are
\be
h_a=\bar h_a=\frac12a(1-a)=\frac12\lt(\frac{k}{n}+\frac{\m}{2\pi n}\rt)\lt(1-\frac{k}{n}-\frac{\m}{2\pi n}\rt)
\ee
and the scaling dimension is
\be
\D_a=h_a+\bar h_a=\lt(\frac{k}{n}+\frac{\m}{2\pi n}\rt)\lt(1-\frac{k}{n}-\frac{\m}{2\pi n}\rt).
\ee
The partition function satisfies the following differential equation $(z_{ij}\equiv z_i-z_j)$
\be
\p_{z_2}\ln Z_{\mathrm{qu}}(z_i,\bar z_i)=-2h_a\lt(\frac{1}{z_{21}}+\frac{1}{z_{23}}-\frac{1}{z_{24}}\rt)+\frac{A(z_i,\bar z_i)}{z_{21}z_{23}z_{24}}.
\ee
It's convenient to use $SL(2,\mathbb{C})$ symmetry $z\rightarrow\frac{(z_1-z)(z_3-z_4)}{(z_1-z_3)(z-z_4)}$ to fix the locations of three of the four twist operators to $z_1=0,z_2=x\equiv\frac{z_{12}z_{34}}{z_{13}z_{24}},z_3=1,z_4\rightarrow\inf$.
\be
\p_x\ln Z_{\mathrm{qu}}(x,\bar x)=-2h_a\lt(\frac{1}{x}-\frac{1}{1-x}\rt)-\frac{A(x,\bar x)}{x(1-x)}.
\ee
We can use the global monodromy conditions $\ref{Monodromyqu}$ to determine $A(x,\bar x)$. Before doing this, we must introduce the auxiliary correlation function
\be
h(\bar z,w;z_i)=\frac{\langle-\frac12\p_{\bar z} X\p_w\bar X\mT_a(z_1)\td{\mT}_a(z_2)\mT_a(z_3)\td{\mT}_a(z_4)\rangle}{\langle\mT_a(z_1)\td{\mT}_a(z_2)\mT_a(z_3)\td{\mT}_a(z_4)\rangle}
=B(z_i,\bar z_i)\bar\o_{1-a}(\bar z)\o_{1-a}(w),
\ee
which is determined up to the constant factor $B(z_i,\bar z_i)$ in the same way as $g$ was. Then we have
\be
\oint_{\mC_i}dz~g(z,w)+\oint_{\mC_i}d\bar z~ h(\bar z,w)=0.
\ee
Dividing by $\o_{1-a}(w)$ and letting $w\rightarrow\inf$
\be\label{Monodromyqu2}
A\oint_{\mC_i}dz~\o_a+B\oint_{\mC_i}d\bar z~\bar\o_{1-a}=-(1-a)\oint_{\mC_i}dz~(z-x)\o_a.
\ee
All the contour integrals appeared in the above equation can be expressed in terms of the hypergeometric function $F(x)\equiv{ _2F_1}(a,1-a,1,x)$ and its derivative
\be
\begin{split}
\oint_{\mC_1}dz~\o_a=2\pi\mathrm{i}e^{-\pi\mathrm{i}a}F(x),\qquad \oint_{\mC_2}dz~\o_a=2\pi\mathrm{i}F(x),\\
\oint_{\mC_1}d\bar z~\bar\o_{1-a}=2\pi\mathrm{i}e^{-\pi\mathrm{i}a}\bar F(\bar x),\qquad\oint_{\mC_2}d\bar z~\bar\o_{1-a}=-2\pi\mathrm{i}\bar F(1-\bar x),\\
-(1-a)\oint_{\mC_1}dz~(z-x)\o_a=2\pi\mathrm{i}e^{-\pi\mathrm{i}a}x(1-x)\frac{dF(x)}{dx},\\
-(1-a)\oint_{\mC_2}dz~(z-x)\o_a=2\pi\mathrm{i}x(1-x)\frac{dF(x)}{dx}.
\end{split}
\ee
Solving eq.~(\ref{Monodromyqu2}) for $A$
\be
A(x,\bar x)=x(1-x)\p_x\ln I(x,\bar x),\quad I(x,\bar x)=F(x)\bar F(1-\bar x)+\bar F(\bar x)F(1-x).
\ee
Then
\be
Z_{\mathrm{qu}}(x,\bar x)=\frac{\mathrm{const}}{|x(1-x)|^{2\D_a}}\frac{1}{I(x,\bar x)}.
\ee
Then to compute the classical action, we need to find properly normalized classical solutions $X_{\mathrm{cl}}(z,\bar z)$ and $X_{\mathrm{cl}}(z,\bar z)$. From the equations of motion, it's easily sees that $\p_z X_{\mathrm{cl}}$ and $\p_z\bar X_{\mathrm{cl}}$
are holomorphic while $\p_{\bar z} X_{\mathrm{cl}}$ and $\p_{\bar z}\bar X_{\mathrm{cl}}$ are antiholomorphic
\be\label{classicalX}
\begin{split}
&\p X_{\mathrm{cl}}(z)=\mathfrak{a}~\o_a(z),\quad \bar\p X_{\mathrm{cl}}(\bar z)=\mathfrak{b}~\bar\o_{1-a}(\bar z),\\
&\p\bar X_{\mathrm{cl}}(z)=\td{\mathfrak{a}}~\o_a(z),\quad \bar\p\bar X_{\mathrm{cl}}(\bar z)=\td{\mathfrak{b}}~\bar\o_{a}(\bar z).
\end{split}
\ee
We can first construct two classical solutions which have simple global monodromy:
\be\label{classicalX1}
\D_{\mC_i}X_{\mathrm{cl},j}=\D_{\mC_i}\bar X_{\mathrm{cl},j}=2\pi\d_{ij},\quad i,j=1,2
\ee
Let $\mathfrak{a}_i,\mathfrak{b}_i,\td{\mathfrak{a}}_i$ and $\td{\mathfrak{b}}_i$ be the coefficients for $X_{\mathrm{cl},i}$ and
$\bar X_{\mathrm{cl},i}$ in eq.~(\ref{classicalX1}), then
\be
\begin{split}
\mathfrak{a}_1=-e^{2\pi\mathrm{i}a}\td{\mathfrak{a}}_1=-\mathrm{i}e^{\pi\mathrm{i}a}\frac{\bar F(1-\bar x)}{I(x,\bar x)},\quad
\mathfrak{a}_2=\td{\mathfrak{a}}_2=-\mathrm{i}\frac{\bar F(\bar x)}{I(x,\bar x)},\\
\mathfrak{b}_1=-e^{2\pi\mathrm{i}a}\td{\mathfrak{b}}_1=-\mathrm{i}e^{\pi\mathrm{i}a}\frac{F(1-x)}{I(x,\bar x)},\quad
\mathfrak{b}_2=\td{\mathfrak{b}}_2=\mathrm{i}\frac{F(x)}{I(x,\bar x)}.
\end{split}
\ee
Therefore the coefficients in eq.~(\ref{classicalX}) reads
\be
\mathfrak{a}=\mathfrak{a}_1\l_1+\mathfrak{a}_2\l_2,\quad \mathfrak{b}=\mathfrak{b}_1\l_1+\mathfrak{b}_2\l_2,\quad \td{\mathfrak{a}}=\td{\mathfrak{a}}_1\bar{\l}_1+\td{\mathfrak{a}}_2\bar{\l}_2,\quad \td{\mathfrak{b}}=\td{\mathfrak{b}}_1\bar{\l}_1+\td{\mathfrak{b}}_2\bar{\l}_2,
\ee
with $\l_{1,2}\in (1-e^{2\mathrm{i}\pi a})\L$.
The classical action is then obtained straightforwardly
\be\label{ClassicalAction}
S_a^{\mathrm{cl}}(\l_1,\l_2)=\pi\sin(\pi a)\lt[\frac{|\tau_a|^2}{\beta_a}|\xi_1|^2+\frac{\a_a}{\beta_a}(\xi_1\bar\xi_2\bar\g+\bar\xi_1\xi_2\g)+\frac{|\xi_2|^2}{\beta_a}\rt].
\ee
where $\g=-\mathrm{i}e^{-\mathrm{i}\pi a}$ and $\xi_j (j=1,2)$ are generic vectors of the target space lattice $\L$.
\section{Normalisation of $\mF_{n}(\m,x)$}\label{AppendixB}
The Siegel theta function $\Theta(\mathbf{0}|M)$ for any symmetric $n\times n$ matrix $M$ with positive imaginary part can be rewritten as
\be
\begin{split}
\Theta(\mathbf{0}|M)=\sum_{\mathbf{m}\in\mathbb{Z}^n}e^{\mathrm{i}\pi\mathbf{m}^{\mathrm{t}}\cdot M\cdot\mathbf{m}}
=\int d^n\mathbf{x}\sum_{\mathbf{m}\in\mathbb{Z}^n}\d^n(\mathbf{x}-\mathbf{m})e^{\mathrm{i}\pi\mathbf{x}^{\mathrm{t}}\cdot M\cdot\mathbf{x}}.
\end{split}
\ee
Using Poisson resummation formula
\be
\sum_{\mathbf{m}\in\mathbb{Z}^n}\d^n(\mathbf{x}-\mathbf{m})=\sum_{\mathbf{p}\in\mathbb{Z}^n}e^{2\pi\mathrm{i}\mathbf{p}\cdot\mathbf{x}},
\ee
we have
\be\label{Theta}
\begin{split}
\Theta(\mathbf{0}|M)=\sum_{\mathbf{p}\in\mathbb{Z}^n}\int d^n\mathbf{x}~e^{\mathrm{i}\pi\mathbf{x}^{\mathrm{t}}\cdot M\cdot\mathbf{x}+2\pi\mathrm{i}\mathbf{p}\cdot\mathbf{x}}
=\sum_{\mathbf{p}\in\mathbb{Z}^n}\frac{1}{\sqrt{\det(-\mathrm{i}M)}}e^{-\mathrm{i}\pi\mathbf{p}^{\mathrm{t}}\cdot M^{-1}\cdot\mathbf{p}}
=\frac{\Theta(\mathbf{0}|-M^{-1})}{\sqrt{\det(-\mathrm{i}M)}}.
\end{split}
\ee
\par Now applying this formula to our problem. Firstly, we have
\be
\eta\td{\O}=-A(\O/\eta)^{-1}A,
\ee
where $A=U^{\dg}HU$ with $H=\diag(\cdots,(-\o_p\td{\o}_p)^{1/2},\cdots)$.
Then by setting $M=\eta\td{\O}$ in eq.~(\ref{Theta}), we find
\be
\Theta(\mathbf{0}|\eta\td{\O})=\frac{\sqrt{\det(\O/(\mathrm{i}\eta))}}{\det A}\Theta(\mathbf{0}|A^{-1}(\O/\eta)A^{-1}).
\ee
 We also find
\be
\frac{\det(\O/(\mathrm{i}\eta))}{(\det A)^2}=\prod_{k=1}^{n}\frac{-\mathrm{i}\o_k}{(-\eta\o_k\td{\o}_k)}=\prod_{k=1}^{n}(-\mathrm{i}\eta\td{\o}_k)^{-1}
\ee
and
\be
[A(\O/\eta)^{-1}A]_{rs}=\frac{\eta}{n}\sum_{p=1}^n\td{\o}_p^{-1}\cos[\frac{2\pi p}{n}(r-s)]
\ee
\par In the limit $x\rightarrow0$, we have $F_a(x)\rightarrow 1$, $\b_a(x)\rightarrow\inf$, thus $\Theta(\mathbf{0}|\eta\O)\rightarrow 1$. Then
\be
\begin{split}
&\lim_{x\rightarrow0}\frac{[\Theta(\mathbf{0}|\eta\O)\Theta(\mathbf{0}|\eta\tilde{\O})]^2}{\prod_{k=0}^{n-1}\beta_a[F_a(x)]^2}=\lim_{x\rightarrow0}\frac{\det(\O/(\mathrm{i}\eta))}{(\det A)^2\prod_{k=0}^{n-1}\beta_{\frac{k}{n}+\frac{\m}{2\pi n}}}
=\eta^{-n}\lim_{x\rightarrow0}\prod_{p=1}^{n}(-\mathrm{i}\td{\o}_p\beta_{\frac{p-1}{n}+\frac{\m}{2\pi n}})^{-1}\\
&=\eta^{-n}\Big[2\sin\lt(\frac{\m}{2n}\rt)\Big]^{-1}\prod_{p=1}^{n-1}\lt[\sin\Big(\frac{\pi p}{n}+\frac{\m}{2n}\Big)+\sin\Big(\frac{\pi p}{n}-\frac{\m}{2n}\Big)\rt]^{-1}\\
&=\frac{1}{2n\eta^n}\Big[\sin\lt(\frac{\m}{2n}\rt)\Big]^{-1}\Big[\cos\Big(\frac{\m}{2n}\Big)\Big]^{1-n},
\end{split}
\ee
where we have used the identity
\be
\prod_{p=1}^{n-1}\lt[\sin\Big(\frac{\pi p}{n}+\frac{\m}{2n}\Big)+\sin\Big(\frac{\pi p}{n}-\frac{\m}{2n}\Big)\rt]=n\Big[\cos\Big(\frac{\m}{2n}\Big)\Big]^{n-1}.
\ee
Thus, the normalisation constant $s_{n,\m}$ of $\mF_n(\m,x)$ is obtained.
\end{appendix}
\bibliography{2021}
\bibliographystyle{ieeetr}
\end{document}